\documentclass[preprint]{aastex}
\slugcomment{accepted for publication in Astrophysical Journal}
\shorttitle{\ion{O}{6} Emission in the Milky Way}
\shortauthors{B. Otte et al.}

\begin{document}
\title{The {\em Far Ultraviolet Spectroscopic Explorer} Survey of \ion{O}{6}
Emission in the Milky Way}
\author{Birgit Otte\altaffilmark{1} and W. Van Dyke Dixon}
\altaffiltext{1}{Current address: Department of Astronomy, University of
Michigan, 500 Church Street, Ann Arbor, MI 48109}
\affil{Department of Physics \& Astronomy, The Johns Hopkins University, 3400
North Charles Street, Baltimore, MD 21218}
\email{otteb@umich.edu, wvd@pha.jhu.edu}


\begin{abstract}
We present a survey of \ion{O}{6} $\lambda$1032 emission in the Milky Way using
data from the {\em Far Ultraviolet Spectroscopic Explorer (FUSE)} satellite. The
observations span the period from launch in 1999 to July 2003. Our survey
contains 112 sight lines, 23 of which show measurable \ion{O}{6} $\lambda$1032
emission. The \ion{O}{6} $\lambda$1032 emission feature was detected at all
latitudes and exhibits intensities of 1900--8600\,photons~s$^{-1}$\,cm$^
{-2}$\,sr$^{-1}$. Combined with values from the literature, these emission
measurements are consistent with the picture derived from recent \ion{O}{6}
absorption surveys: high-latitude sight lines probe \ion{O}{6}-emitting gas in a
clumpy, thick disk or halo, while low-latitude sight lines sample mixing layers
and interfaces in the thin disk of the Galaxy.
\end{abstract}


\section{INTRODUCTION}

Hot interstellar gas cooling through temperatures of a few times $10^5$ K is
most easily studied through observations of the lithium-like ions \ion{C}{4},
\ion{Si}{4}, \ion{N}{5}, and \ion{O}{6}. The \ion{O}{6} resonance transitions at
1031.93 and 1037.62\,\AA\ provide the primary cooling mechanism for
collisionally-ionized gas at temperatures near $3\times10^5$\,K, while the
corresponding lines of \ion{N}{5} ($\lambda\lambda 1238.81, 1242.80$),
\ion{C}{4} ($\lambda\lambda 1548.19, 1550.76$), and \ion{Si}{4} ($\lambda\lambda
1393.76, 1402.77$) dominate at cooler temperatures \citep{sd}. Extensive studies
of interstellar \ion{N}{5}, \ion{C}{4}, and \ion{Si}{4} absorption have been
performed with the {\em International Ultraviolet Explorer (IUE)}
\citep[e.g.,][]{sem92, sem93} and with the GHRS and STIS spectrographs aboard
the {\em Hubble Space Telescope} \citep[e.g.,][]{spi}, but, except for a survey
of \ion{O}{6} absorption toward 72 nearby stars conducted with the {\em
Copernicus} satellite \citep{j78a, j78b}, large surveys of \ion{O}{6} in the
Galactic disk and halo have had to await the launch of {\em FUSE}, the {\em Far
Ultraviolet Spectroscopic Explorer}. Recent \ion{O}{6} absorption-line surveys
include \citet{wak03}, \citet{sav}, \citet{bowen}, and \citet{sav06}.
\citet{inda, indb} recently obtained both {\em FUSE} and {\em HST} observations
toward 34 stars and compared the derived ionization ratios of the Li-like
absorbers with various theoretical models.

By themselves, absorption lines yield the total column of the absorbing species
integrated along the line of sight, but when combined with emission-line data,
they can be used to derive the local density of the emitting gas \citep{ss},
providing a powerful diagnostic of the microphysics of cooling in these regions.
\ion{O}{6} emission has been observed with {\em FUSE} along a handful of sight
lines, but more observations are required to investigate the distribution,
kinematics, and emission mechanisms of \ion{O}{6}-bearing gas in the Galaxy. We
searched the {\em FUSE} archive for additional sight lines with \ion{O}{6}
emission. From a sample of 112 sight lines, we detect \ion{O}{6} emission in 23.
We describe our data sample and reduction techniques in \S\S\ \ref{sam} and
\ref{red}. Section \ref{mupl} explains the data fitting and presents our
results. In \S\ \ref{dis}, we compare these results with various models and 
with observations obtained at other wavelengths and discuss our results. We
summarize our findings in \S\ \ref{conc}.

\section{\label{sam}OBSERVATIONS}

The {\em FUSE}\/ instrument consists of four independent optical paths. Two
employ LiF-coated gratings and mirrors that are sensitive to the wavelengths
between 990 and 1187\,\AA, and two use SiC coatings that are sensitive to
wavelengths between 905 and 1100\,\AA. Each spectrograph possesses three
apertures that simultaneously observe different parts of the sky. The low
resolution (LWRS) aperture samples an area of $30\arcsec\times30\arcsec$. The
medium resolution (MDRS) aperture spans $4\arcsec\times20\arcsec$ and lies about
$3\farcm5$ from the LWRS aperture. The high resolution (HIRS) aperture lies
between the MDRS and LWRS apertures (about $1\farcm8$ from the LWRS aperture)
and samples an area $1\farcs25\times20\arcsec$ wide. A complete description of
{\em FUSE}\/ can be found in \citet{moos} and \citet{sahn}.

The data used in our survey consist of LiF1A LWRS spectra from the S405 and Z907
programs (as of 2003 July 25) taken in time-tag mode. The S405 programs
represent background observations and observations for alignment of the four
optical paths. The Z907 programs contain observations of active galactic nuclei
(AGNs) and other extragalactic targets that can be used as background sources
for absorption studies. In some cases, these targets were too faint for use as
background sources, which resulted in relatively flat continua suited for the
search of Galactic \ion{O}{6} emission. In addition to these two programs, we
searched the Multimission Archive at Space Telescope (MAST; as of 2003 July 25)
for MDRS observations of point sources obtained in time-tag mode, as their LWRS
apertures should sample only background radiation. If HIRS time-tag observations
exist in addition to the MDRS observations of a target, the corresponding LWRS
spectra of both MDRS and HIRS observations were combined into one spectrum. When
combining LWRS data from HIRS and MDRS observations obtained at different
epochs, we ignored the fact that the LWRS aperture may sample slightly different
regions of the sky depending on the aperture and position angle used for the
observations. At worst, this simplification decreases the spatial resolution of
our data from 30 arcsec (the width of the LWRS aperture) to 7 arcmin (twice the
distance between LWRS and MDRS aperture).

The sample was further modified as follows: All extracted LWRS spectra that
showed continuum flux were removed from the survey. We also removed sight lines
whose \ion{O}{6} emission could be identified as originating in the supernova
remnants Vela or the Cygnus Loop, since these structures do not represent the
diffuse interstellar medium. Sight lines passing within a few arc minutes of the
hot white dwarf KPD~0005+511 exhibit unusually strong \ion{O}{6} emission (up to
25,000\,LU). We determined that the emission from this region originates in a
high-ionization planetary nebula (HIPN) around the white dwarf (Otte, Dixon, \&
Sankrit 2004), and therefore removed all observations toward this object from
the survey.

Three additional white dwarfs exhibit highly ionized circumstellar environments
possibly probed by our survey; \ion{Si}{4}, \ion{C}{4}, and \ion{N}{5} have been
detected in their {\em IUE} and/or STIS spectra \citep[Holberg, Barstow, \& Sion
1998;][]{bann}. One of these stars, the white dwarf WD~0455-282, lies within a
few arc minutes of our sight lines P10411 ($l=229\fdg3$) and S40537
($l=230\fdg7$), both of which exhibit \ion{O}{6} emission. \citet{holb} quote
absorption in {\em IUE} data at $v_{\rm helio}=16\pm3$\,km\,s$^{-1}$ and suggest
that this absorption is circumstellar. The \ion{O}{6} emission of S40537
($v_{\rm helio}=-7\pm19$\,km\,s$^{-1}$) is less likely to be of circumstellar
origin than the emission observed along sight line P10411
($v_{\rm helio}=+10\pm12$\,km\,s$^{-1}$). The low temperature of WD~0455-282
\citep[$T_{\rm eff}=57,200$\,K;][]{sav06}, however, makes the photoionization of
the circumstellar material to high ionization levels such as \ion{O}{6}
difficult. While we cannot rule out the possibility of a HIPN around
WD~0455-282, we include P10411 in our sample of diffuse \ion{O}{6} emission. The
nearby LWRS spectra of the other two white dwarfs, KPD~1034+001 (S40509,
$l=247\fdg6$) and G191--B2B (S40576, $l=156\fdg0$), do not show any measurable
\ion{O}{6} emission.

Five of our sight lines lie within 5\degr\ of the ecliptic plane, where solar
\ion{O}{6} emission scattered off interplanetary dust is a concern, but we have
found no evidence for such contamination in the data. P12012 ($l=194\fdg3$) is a
science observation with Jupiter in the MDRS aperture. Its LiF1A spectrum shows
\ion{O}{6} emission with a heliocentric velocity consistent with zero. Its
night-only SiC2A spectrum, however, shows no \ion{C}{3} $\lambda$977 emission,
which would be bright if the \ion{O}{6} emission were solar. The atmosphere of
Jupiter consists primarily of H$_2$. While there is a possibility that the
observed emission comes from H$_2$ fluorescence excited by solar \ion{O}{6}
emission, the absence of other H$_2$ emission lines and the 3\farcm5 separation
between the MDRS and the LWRS aperture argue against this possibility. The other
four sight lines in the ecliptic plane show no \ion{O}{6} emission.

Sight line P10413 ($l=162\fdg6$) is a science observation with Capella in the
MDRS aperture. The stellar spectrum shows strong \ion{O}{6} emission lines with
a possible contamination of the HIRS and LWRS aperture. The measured velocity of
the \ion{O}{6} emission line in the LWRS aperture, however, is inconsistent with
the radial velocities of either star in this spectroscopic binary during the
orbital phase at the time of the {\em FUSE} observations \citep{young}.

\begin{figure}
\plotone{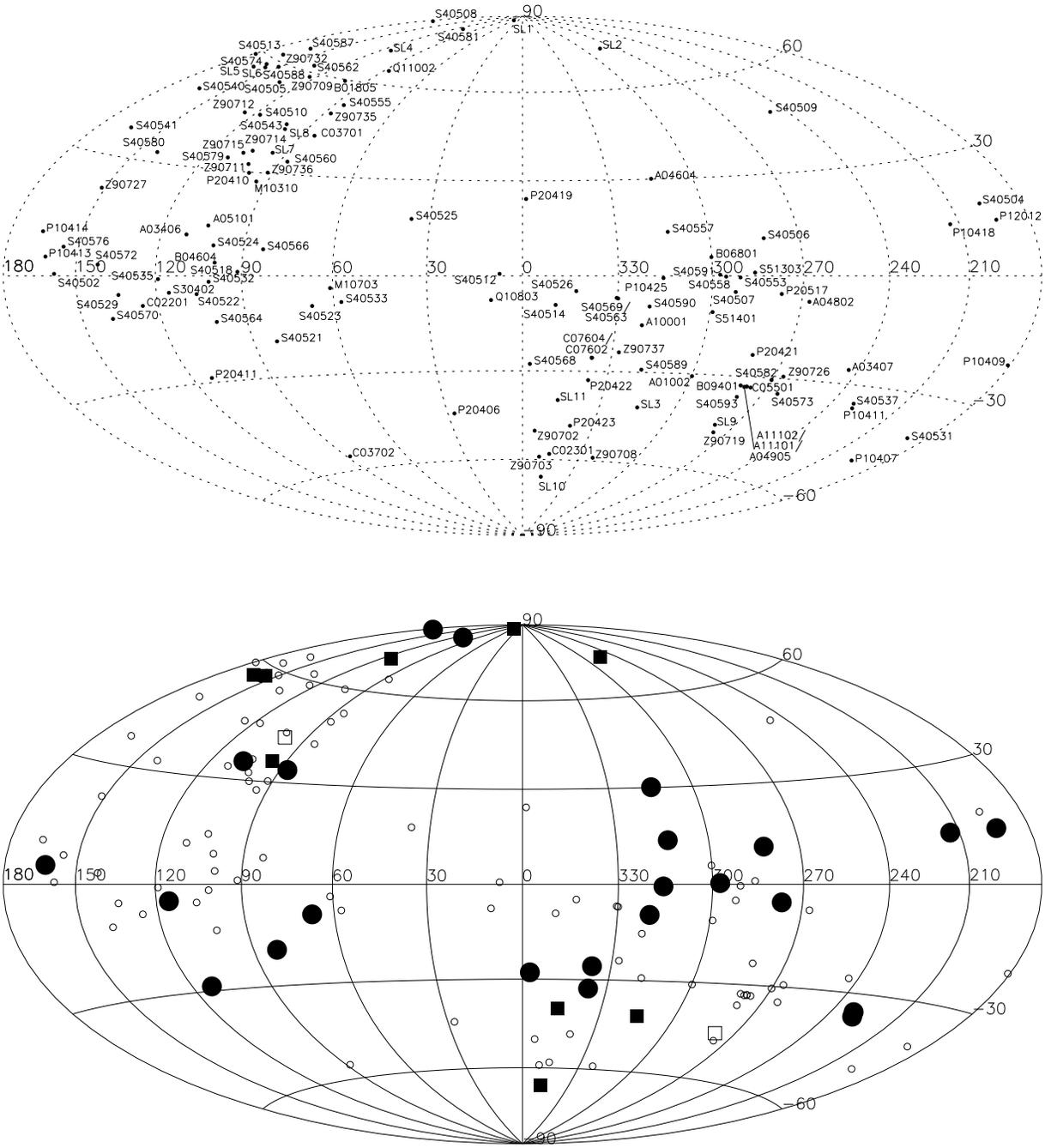}
\caption{\label{pos} Distribution of the sight lines in the \ion{O}{6} emission
survey. In the top panel, each sight line is labeled by its identifier used in
Tables \ref{det}--\ref{pub}. Galactic coordinates are used in Hammer-Aitoff
projection. In the bottom panel, filled symbols represent \ion{O}{6}
$\lambda$1032 detections; open symbols show non-detection sight lines. Circles
mark the survey sight lines, squares represent previously published \ion{O}{6}
emission sight lines. The circles for upper limit survey sight lines were
reduced in size to avoid overcrowding the graph.}
\end{figure}

In a few cases, we combined into a single spectrum sight lines that are
separated by less than $0\fdg1$ to improve the signal-to-noise ratio. Table
\ref{obsid} lists the target IDs (usually the first 6--8 characters of the
observation ID) by which we identify the sight lines in this survey and the
observations that contribute to each. The original sample contained 225 sight
lines, of which 77 were removed because of continuum flux or even absorption
features in the LWRS aperture; 16 more sight lines were removed because of SNR
or HIPN emission. Another 16 sight lines were combined with nearby sight lines,
and four observations turned out not to contain any night time data. The final
survey contains 112 sight lines unevenly distributed on the sky (Fig.\
\ref{pos}). The apparent clustering of observations in two quadrants can be
explained by observational constraints that combine to reduce target
availability around the orbital plane, which stretches across the Galactic
center. 

\section{\label{red}DATA REDUCTION}

In several cases, observations of the same target were repeated months or even
years apart. Using the CalFUSE pipeline to combine these time-tag observations
can yield faulty wavelength and thus velocity information. We developed software
in IDL\footnote{Interactive Data Language; IDL is a registered trademark of
Research Systems, Inc. for their Interactive Data Language software.} to
optimize the extraction of the \ion{O}{6} emission from the LiF1A channel by
calibrating each exposure individually in wavelength (using \ion{O}{1} airglow
lines) and flux before combining them into one spectrum. We used the raw data
and applied the first steps of the CalFUSE pipline (version 2.3; initializing
the header, computing Doppler corrections). Our IDL program then extracted all
photons that were detected in a rectangular area around the expected location of
\ion{O}{6} in the LWRS spectrum. This rectangular region was determined from the
data set S4054801, a very long background observation.

The LWRS spectrum was collapsed perpendicular to the wavelength dispersion axis
to create a one-dimensional spectrum. The S4054801 data yielded the template
positions of the five \ion{O}{1} airglow lines ($\lambda\lambda$1027.431,
1028.157, 1039.230, 1040.943, and 1041.688) surrounding the \ion{O}{6}
$\lambda\lambda$1032,1038 doublet. The strongest of these airglow lines is
\ion{O}{1} $\lambda$1027. Our IDL program searched for an emission line in each
individual exposure in a 160 pixel wide window around the \ion{O}{1}
$\lambda$1027 template position and fitted its position with a Gaussian profile.

In the next step, the program used the first fitted position and the template
offsets between the airglow lines to search for the remaining four \ion{O}{1}
emission lines and to fit their positions. The airglow fits were then compared
with empirical criteria for good airglow fits. The three criteria are (1) The
ratio of the fitted peak intensity over the continuum noise should be $>0.6$.
(2) The fitted Gaussian line width should be $>10$ and $<50$ pixels. (3) The
fitted position of each airglow line should be less than 15 pixels from its
expected position relative to the measured \ion{O}{1} $\lambda$1027 line.
Airglow lines not fulfilling all three criteria were marked as bad and not
considered further.

An intensity-weighted average offset from the template positions was calculated
for each exposure. For exposures without any good airglow lines, average offsets
were determined by interpolation between the measured offsets of the exposures
closest in time before and after the exposure in question.

The next step was to determine the correction factor for the time-dependent
effective area for each exposure and to shift each exposure according to its
average offset from the template. The spectra were then combined, and the
airglow lines were fitted once more in the combined, higher signal-to-noise
spectrum. The airglow line fits were again compared with the criteria for good
fits. The good fits were used to determine the geocentric wavelength solution,
which was assumed to be linear in the \ion{O}{6} emission region. The
differences between the standard CalFUSE wavelength calibration and our linear
wavelength solutions are less than 0.02\,{\AA} around $\lambda=1032$\,\AA. If no
good airglow lines were found in the combined spectrum, the template positions
were used. Shifts in wavelength caused by grating or mirror motions on orbital
timescales are not corrected in this case. The only five sight lines without
good airglow lines were non-detections and therefore did not require additional
correction. If only one good airglow line was found, it was used to determine an
offset from the template positions and their wavelength solution. All necessary
correction factors, shifts, and wavelength solutions were saved, so that they
could be applied in the second and final extraction of the spectra.

For each sight line, a day-plus-night and a night-only spectrum were finally
extracted. To create these spectra, the uncorrected exposures were used to
extract the photons of the LiF1A LWRS aperture according to their day/night
flag. Pulse height limits 2--25 were used, since this range appears suited for
all observations taken between 1999 (launch) and 2003 (start of our survey
work). The extracted region of detector segment 1A was collapsed perpendicular
to the dispersion axis to create the one-dimensional day-plus-night and
night-only spectrum. Both the day-plus-night and the night-only spectra of each
exposure were shifted by the previously determined offset and adjusted to a
heliocentric wavelength scale for each spectrum. The spectra were corrected for
the time-dependent effective area by the factor calculated previously, and then
combined into one day-plus-night and one night-only spectrum for each sight
line. The one-dimensional spectra were binned by 16 pixels ($\sim0.108$\,\AA)
and saved as ASCII tables. Further data analysis was performed on the night-only
spectra to avoid contamination of the \ion{O}{6} measurements with possible
airglow emission; the day-plus-night spectra were used only to assist, e.g., in
the identification of possible continuum absorption features.

\section{\label{mupl}\ion{O}{6} MEASUREMENTS}

We distinguished between \ion{O}{6} detections and non-detections solely by the
spectrum's appearance, i.e., no automated detection algorithm was used. Thus,
our \ion{O}{6} measurements include mostly sight lines with at least 2$\sigma$
detections. Each detection spectrum was fitted twice using the
IRAF\footnote{IRAF is distributed by the National Optical Astronomy
Observatories, which are operated by the Association of Universities for
Research in Astronomy, Inc., under cooperative agreement with the National
Science Foundation.} routine SPECFIT \citep{kriss}. Before fitting the data,
we smoothed the error array slightly to avoid pixels with a flux error of zero.
If the average continuum count per pixel in the measurement region was too low
(i.e., $<4$\,cts/pxl), we binned the spectrum by an additional factor of two. We
first calculated the reduced $\chi_{\rm c}^2/\nu_{\rm c}$ value for a linear
continuum fit to the entire measurement region from 1028.5 to 1036.5\,\AA. The
second fit (yielding $\chi^2/\nu$) included an emission line model consisting of
a convolution of a Gaussian profile with a 106\,km\,s$^{-1}$ wide flat-top
profile (representing the LWRS aperture).

\begin{figure}
\plotone{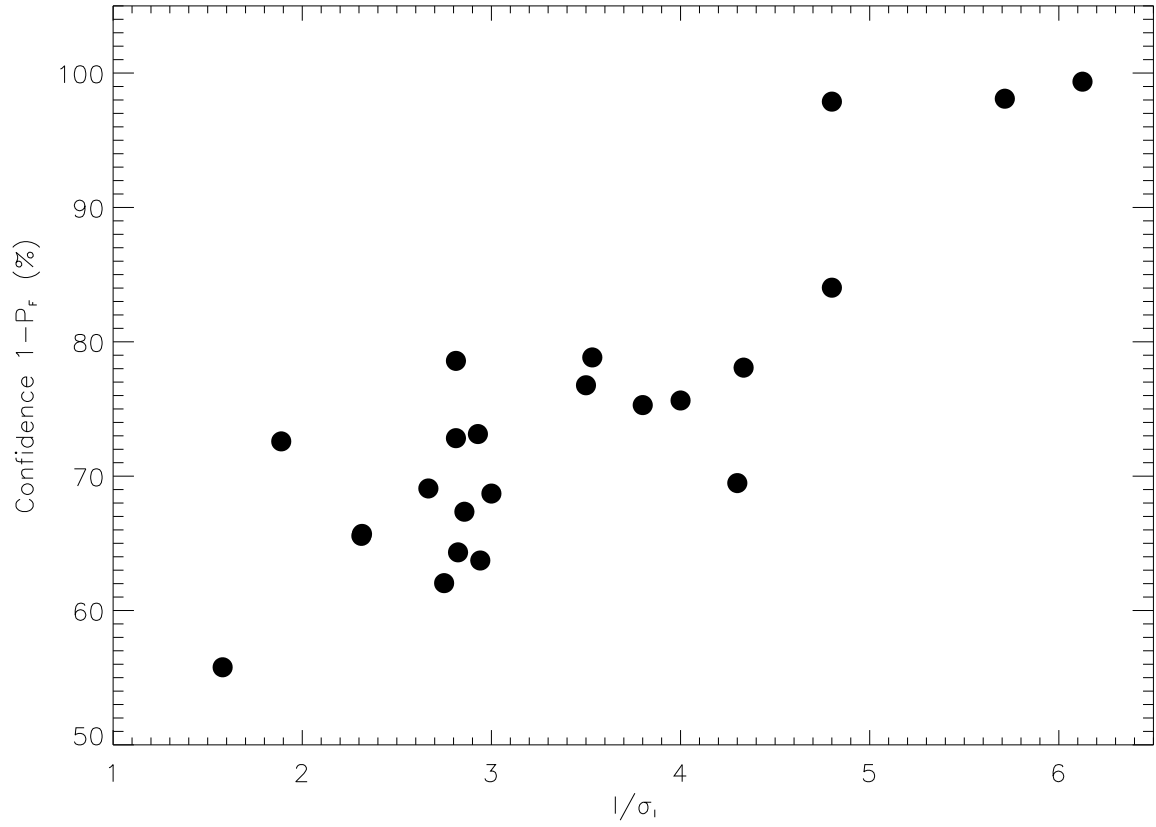}
\caption{\label{ft} Comparison between the confidence level and the
signal-to-noise ratio. The confidence level $1-P_{\rm F}$ was obtained from the
F test using the ratio of the variances of the continuum fit and the emission
line fit for each detection. The signal-to-noise ratio $I/\sigma_{\rm I}$ was
derived from the actual emission line measurements.}
\end{figure}

We used the routine MPFTEST from the Markwardt IDL library to perform the
standard F test ($F=[\chi_{\rm c}^2/\nu_{\rm c}]/[\chi^2/\nu]$) on each
detection spectrum to obtain a confidence level for the existence of the
\ion{O}{6} $\lambda$1032 emission line. Figure \ref{ft} compares the confidence
level $1-P_{\rm F}$ with the ratio of the measured intensity to its uncertainty
$I/\sigma_{\rm I}$. Only three sight lines have a signal-to-noise ratio
significantly lower than 3.0 (i.e., less than 2.5). Only one of these sight
lines also has a confidence level below 60\% (S40523). Since the emission line
along sight line S40523 is stronger than any possible emission feature in our
non-detection sight lines, we kept S40523 as a detection, but with the caveat of
its higher uncertainty. The analysis was performed on detections with
$I/\sigma_{\rm I}\geq2.7$, which corresponds to a 99.3\% confidence level,
unless noted otherwise. (Including the lower signal-to-noise detections would
change the quoted means, uncertainties, and standard deviations in the
subsequent paragraphs by only 100\,LU.)

Three-sigma upper limits for \ion{O}{6} $\lambda$1032 non-detections were
derived from the standard deviation in the continuum counts, which we will
discuss in more detail in \S\ \ref{dlim}. No background subtraction was applied,
i.e., the continuum includes any detector background and its uncertainty.
Following the arguments of \citet{sh01}, we were able to exclude contamination
of the \ion{O}{6} detections by stray light or scattered solar \ion{O}{6}
emission. Both emission and absorption lines of molecular hydrogen can change
the appearance of the \ion{O}{6} $\lambda$1032 emission line. The two closest
H$_2$ absorption lines are from the transitions Lyman (6,0) $P(3)$
$\lambda$1031.19 and $R(4)$ $\lambda$1032.35. We believe, however, that their
effect on \ion{O}{6} $\lambda$1032 is negligible, as these are weak transitions
in the cold phase of the interstellar medium \citep{sh00}. Fluorescent H$_2$
emission may increase the observed \ion{O}{6} $\lambda$1032 emission. Any H$_2$
emission feature near the \ion{O}{6} doublet is only about 70\% as strong as the
transitions Werner (0,0) $Q(1)$ $\lambda$1009.77 and Werner (0,1) $P(3)$
$\lambda$1058.82 \citep{sh01}. Neither one was observed in any of our sight
lines, thus making H$_2$ contamination of the \ion{O}{6} $\lambda$1032 emission
negligible. The \ion{O}{6} $\lambda$1038 emission line was not fitted, because
it is fainter and thus more difficult to measure than the \ion{O}{6}
$\lambda$1032 line and often blended with the \ion{C}{2}$^\ast$ $\lambda$1037
or \ion{C}{2} $\lambda$1036 emission line.

\begin{figure}
\epsscale{.80}
\plotone{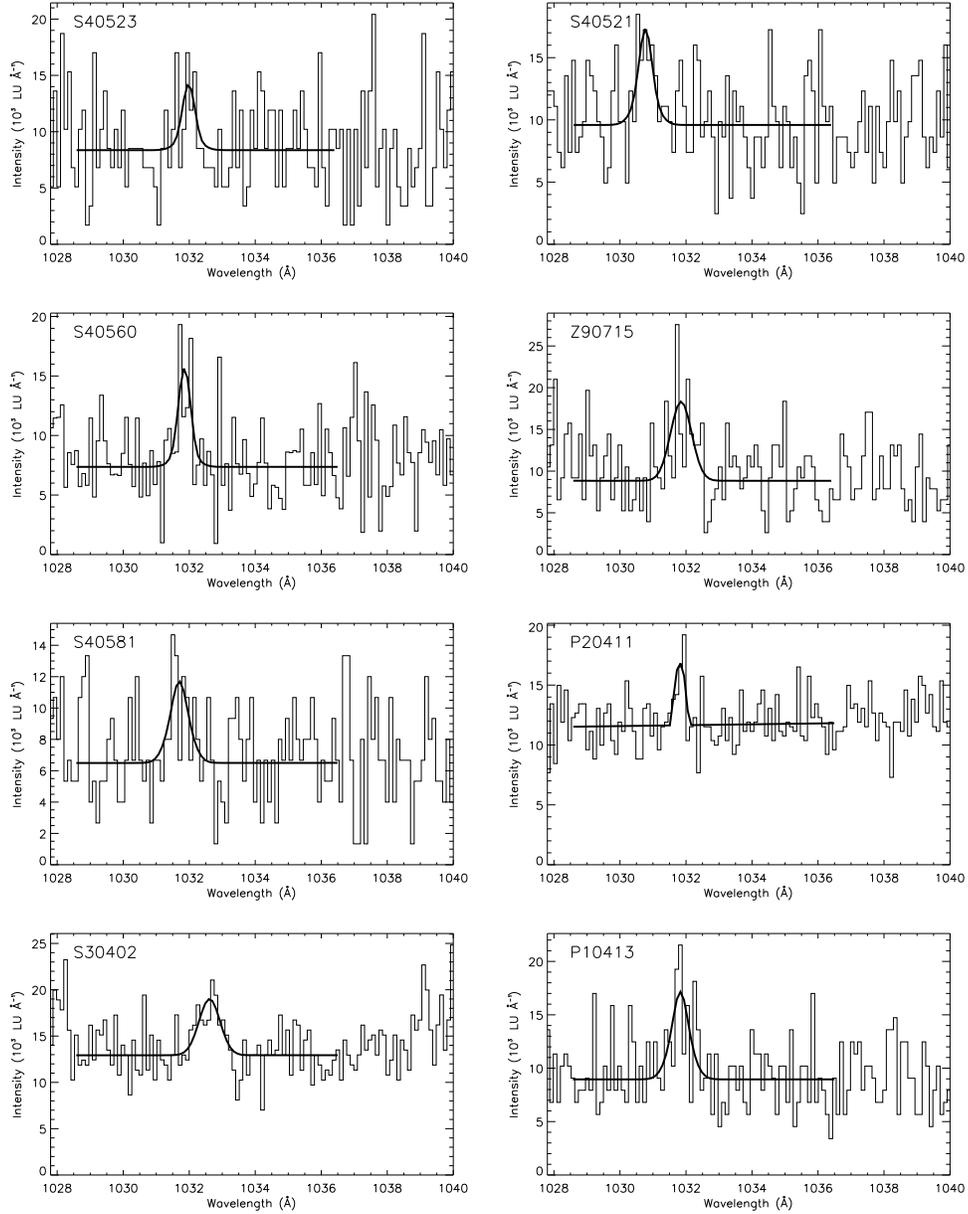}
\caption{\label{detsp} {\em FUSE} \ion{O}{6} emission spectra. The 23 detection
sight lines are plotted in the same order as listed in Table \ref{det}. The
least $\chi^2$ fits are overplotted. The weaker \ion{O}{6} $\lambda$1038
emission line or the \ion{C}{2}$^\ast$ $\lambda$1037 emission line are visible
in a few of the spectra, but were not fitted. Sight lines with longitudes a)
$l=0\degr-170\degr$, b) $l=170\degr-300\degr$, c) $l=300\degr-360\degr$.}
\end{figure}

\begin{figure}
\epsscale{.80}
\plotone{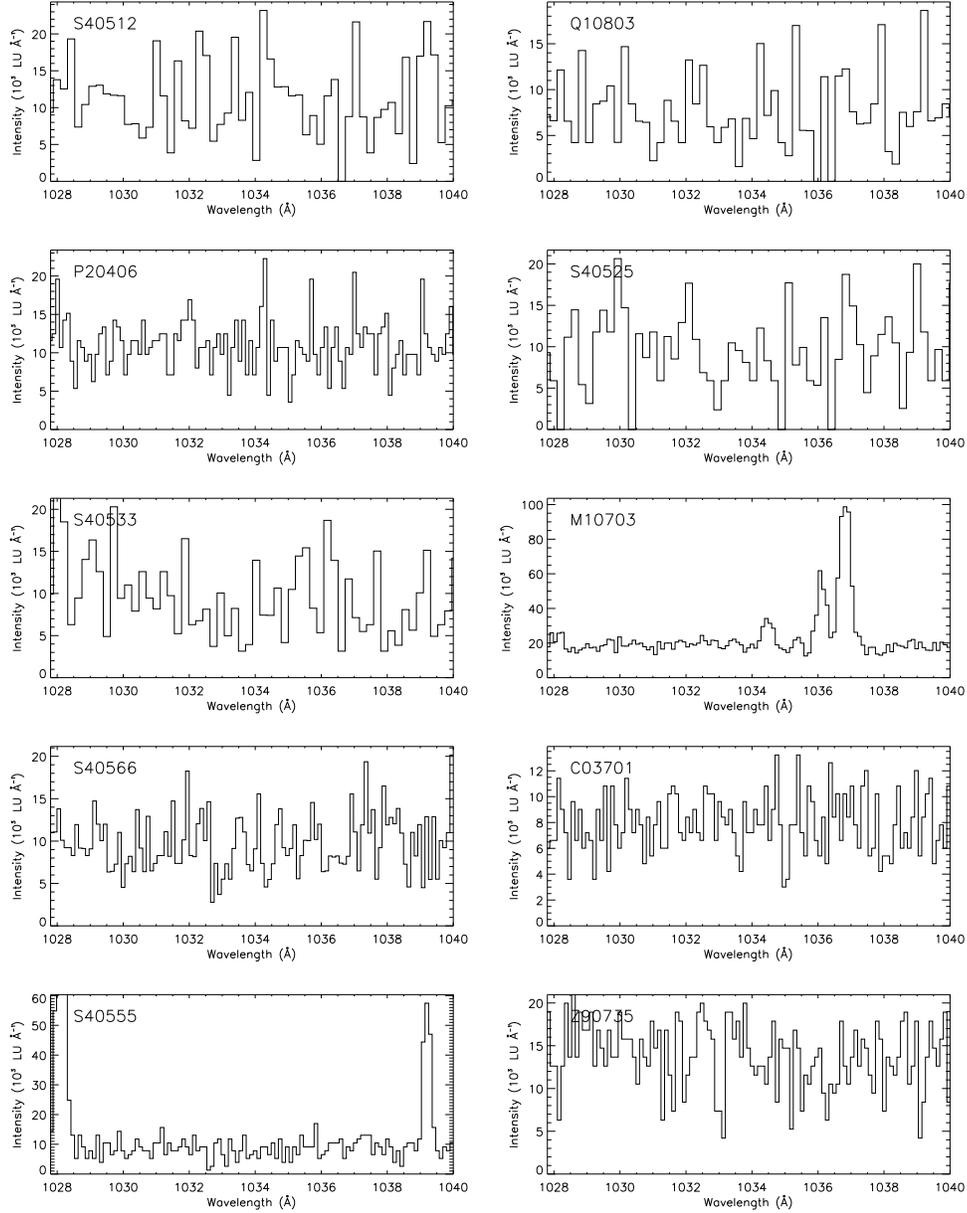}
\caption{\label{uplsp} {\em FUSE} spectra without measurable \ion{O}{6}
emission. The 89 non-detection sight lines are plotted in the same order as
listed in Table \ref{upl}. Sight lines with longitudes a) $l=0\degr-86.92\degr$,
b) $l=86.92\degr-100.8\degr$, c) $l=100.8\degr-120\degr$, d)
$l=120\degr-151\degr$, e) $l=151\degr-180\degr$, f) $l=180\degr-271\degr$, g)
$l=271\degr-291\degr$, h) $l=291\degr-330\degr$, i) $l=330\degr-360\degr$.
Figures 4 is available in its entirety in the electronic edition of the
Astrophysical Journal. Figure 4a is shown here for guidance regarding its form
and content.}
\end{figure}

The results are listed in Tables \ref{det} (detections) and \ref{upl}
(non-detections). The sight lines are sorted by increasing Galactic longitude.
Both tables list the target IDs, the Galactic coordinates $l$ and $b$, and the
total night time $t$. Table \ref{det} also lists the emission line fit
parameters: \ion{O}{6} $\lambda$1032 intensity $I_{\rm 1032}$, the central
wavelength $\lambda_{\rm c}$, the Gaussian FWHM, the reduced $\chi^2$ of the
fit, the confidence level obtained from the F test $1-P_{\rm F}$, and the
derived local-standard-of-rest velocity $v_{\rm LSR}$. Table \ref{upl} lists the
$3\sigma$ upper limit for the \ion{O}{6} $\lambda$1032 intensity. Both tables
include several values from the literature: the total reddening along the line
of sight $E(B-V)$ (inferred from {\em IRAS} 100\,$\mu$m emission; Schlegel,
Finkbeiner, \& Davis 1998), the {\em ROSAT} 1/4\,keV soft X-ray (SXR)
flux\footnote{The SXR flux was determined using the online tool at
heasarc.gsfc.nasa.gov/cgi-bin/Tools/xraybg/xraybg.pl with a 0.4 degree cone
radius.}, the H$\alpha$ intensity $I_{{\rm H}\alpha}$, and a note if the sight
line intersects an \ion{H}{1} high-velocity cloud (HVC). Measurements from the
literature are compared with the \ion{O}{6} measurements in the following
sections. Figure \ref{detsp} shows the night-only spectra for our 23 detection
sight lines. Figure \ref{uplsp} shows the first ten non-detection sight lines;
the complete figure is available electronically.

Table \ref{det} lists only the statistical uncertainties in our measurements.
The uncertainty in the size of the LWRS aperture and the sensitivity of the
instrument add a systematic uncertainty of about 14\% \citep{sh01} to the
numbers in Table \ref{det}. We checked the detections for possible wavelength
shifts caused by grating/mirror motion between day and night observations. The
measured change in the position of the Lyman $\beta$ line is between $-0.02$ and
$+0.04$\,\AA, with 20 of the 23 sight lines showing a shift of
$<\pm0.03$\,\AA\ which corresponds to $<\pm9$\,km\,s$^{-1}$ for \ion{O}{6}
$\lambda$1032. These wavelength shifts are within the uncertainties of the
measured \ion{O}{6} $\lambda$1032 wavelengths, i.e., we cannot distinguish
whether a true mirror/grating motion occured between orbital day and night.

The measured intensities can be corrected for extinction using the $E(B-V)$
values derived by \citet{sfd} and the extinction parameterization of
\citet{fitz} for a $R=3.1$ model. The resulting attenuation along all sight
lines is at least 10\%. We will discuss extinction in more detail in \S\
\ref{dis}. Resonance scattering within the \ion{O}{6}-bearing gas can also have
a strong effect on the intensities. For optically thin gas, the intensity ratio
$I_{1032}/I_{1038}=2$, whereas optically thick plasma yields a ratio of unity.
Absorption by \ion{C}{2} and H$_2$ can modify the intensity ratio even further.
Since the \ion{O}{6} $\lambda$1038 line is generally difficult to measure (as
mentioned above) and its intensity has a large uncertainty, the resulting line
ratio is usually inconclusive. We therefore did not try to measure \ion{O}{6}
$\lambda$1038 and estimate the amount of self-absorption for the sight lines
that show \ion{O}{6} $\lambda$1038 emission.

\ion{O}{6} $\lambda$1032 emission has been observed along several sight lines by
other authors. Table \ref{pub} summarizes these measurements. The sight lines
are listed chronologically by date of publication and are numbered
consecutively. The columns are similar to those in Tables \ref{det} and
\ref{upl}: target ID, Galactic coordinates, \ion{O}{6} $\lambda$1032 intensity
(note: upper limits are 2$\sigma$), Gaussian FWHM, derived
local-standard-of-rest velocity, total reddening along line of sight, SXR flux,
H$\alpha$ intensity, note regarding intersected HVCs, and the reference for the
\ion{O}{6} measurements. With one exception, previously-measured \ion{O}{6}
$\lambda$1032 intensities vary between 2000 and 3300\,LU (line unit; 1\,LU =
1~photon~s$^{-1}$\,cm$^{-2}$\,sr$^{-1}$). The \ion{O}{6} measurements in our
survey span a broader range of intensities, from 1900 to 8600\,LU. Due to the
generally shorter exposure times of the survey data, dectections are biased
toward regions of stronger \ion{O}{6} emission. The average intensity for all
diffuse \ion{O}{6} detections reported to date is $4000\pm300$\,LU.

\subsection{\label{dlim}The Detection Limit of {\em FUSE}}

\begin{figure}
\plotone{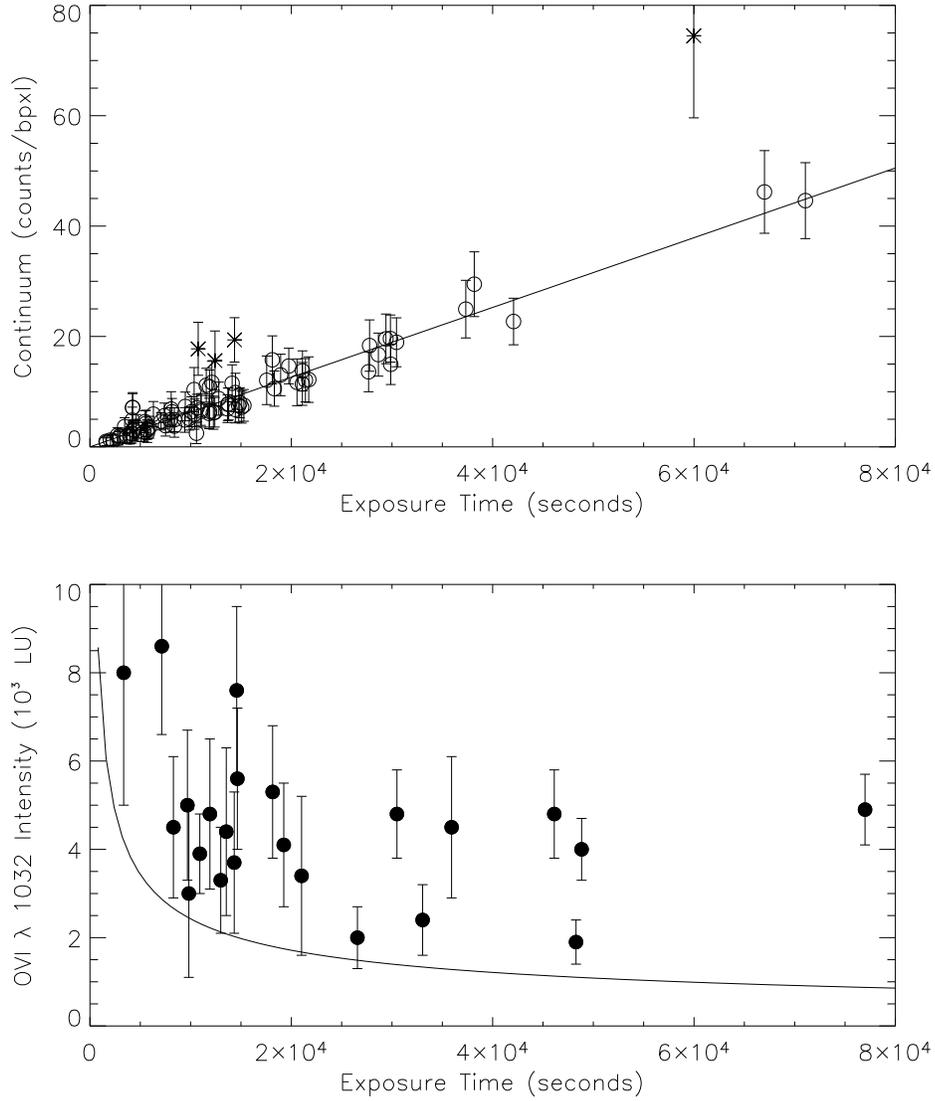}
\caption{\label{uplfit}Relationship between \ion{O}{6} upper limits and exposure
time. Top panel: The average continuum count per binned pixel (between 1030 and
1035\,\AA) is compared with the corresponding exposure time for the
non-detection sight lines. Four data points, which are marked as asterisks, were
not included in the least $\chi^2$ fit (solid line). Bottom panel: The solid
line represents the 3$\sigma$ upper limit for \ion{O}{6} $\lambda$1032 emission
derived from the fit to the average continuum count per binned pixel. The 23
\ion{O}{6} detections from the survey are also shown.}
\end{figure}

Of the 112 sight lines in our survey, 23 exhibit detectable \ion{O}{6} emission.
Intensities range from 1900 up to 8600\,LU with an average intensity of
$4500\pm400$\,LU and a standard deviation (SD) of 1700\,LU around the mean for
individual sight lines. Inclusion of the previously published intensities (Table
\ref{pub}) yields an average of $4000\pm300$\,LU and a SD of 1700\,LU. This
raises the question: How faint an emission line can be detected with
{\em FUSE}\/? For each non-detection sight line, we calculated the average
continuum count per binned pixel between 1030 and 1035\,\AA. The top panel in
Fig.\ \ref{uplfit} reveals a linear relationship between the continuum counts
and the corresponding night exposure time $t$. Four sight lines (marked as
asterisks) have higher than normal continua and were excluded from the least
$\chi^2$ fit (solid line in Fig.\ \ref{uplfit}). We determined the SD $\sigma$
around the continuum using the same region of the spectrum ($1030-1035$\,\AA).
The values of $\sigma$ are comparable to those of the square root of the average
continuum count. It is therefore accurate to convert the fit of the continuum
counts into an analytical expression for the 3$\sigma$ upper limit of the
intensity $I_{1032}$:
$$I_{1032}=\frac{3\,\sqrt{6.32\times10^{-4}\:t}\,\sqrt{54/16}}{t\,A_{\rm
eff}\,\Omega}$$
with $t$ being the night exposure time in seconds, $A_{\rm eff}$ the effective
area, and $\Omega$ the solid angle of the aperture. The uncertainty of the
fitted slope is less than 1.4\% ($9\times10^{-6}$). The factor $\sqrt{54/16}$ is
necessary to scale the continuum counts of one binned pixel (= 16 unbinned
pixels) to the width of the LWRS aperture (54 unbinned pixels) assuming an
unresolved \ion{O}{6} $\lambda$1032 emission line. The 3$\sigma$ upper limit
curve is plotted in the bottom panel of Fig.\ \ref{uplfit} together with all 23
detected \ion{O}{6} intensities. The graph shows that the 3$\sigma$ detection of
intensities between 1000 and 2000\,LU requires very long night exposure times
and that it is almost impossible to detect \ion{O}{6} emission below 1000\,LU at
the 3$\sigma$ confidence level, which is consistent with all of the \ion{O}{6}
detections in Tables \ref{det} and \ref{pub}. The upper limits listed for the
non-detection sight lines in Table \ref{upl} were calculated using the equation
above, but replacing the square root of the fitted continuum counts with the
previously calculated $\sigma$.

\section{\label{dis}DISCUSSION}

\ion{O}{6} absorption is detected in the spectra of UV-bright stars, QSOs, and
AGNs \citep[e.g.,][]{wak03,sav}. Measurements along sight lines through the
Galactic halo indicate that hot, \ion{O}{6}-bearing gas is roughly co-spatial
with the thick disk, having a scale height of about 3.5\,kpc \citep{bowen}.
Within the thick disk, the distribution of \ion{O}{6} is patchy and varies on
small angular scales \citep[$0\fdg05-5\fdg0$ toward the Magellanic Clouds;][]
{howk}. Measurements towards stars in the disk indicate that the \ion{O}{6} is
extremely clumpy: the hot gas cannot exist in uniform clouds \citep{bowen}. The
observations are consistent with the \ion{O}{6} existing in interfaces
\citep{sav06}.

Observation of \ion{O}{6} emission poses different challenges than absorption
observations. There is usually no upper limit on the distance along the line of
sight as is the case with background stars for absorption line studies. The
\ion{O}{6} emission and therefore the signal-to-noise ratio is rather weak
requiring stronger binning of the data and therefore loss of resolution. On the
other hand, no background sources are required, allowing us to search for
emission in any direction of the sky. The picture of the \ion{O}{6} emitting gas
that arises in the following paragraphs is consistent with that derived from the
\ion{O}{6} absorption studies: the \ion{O}{6} bearing gas forms a thick disk
with emission originating most likely from interface regions in the thin disk
and from cooling gas at higher latitudes.

\subsection{\label{p10411}Combining Emission and Absorption Measurements}

Measurements of both absorption and emission along a given line of sight provide
valuable diagnostics. The absorption is proportional to the density of the gas,
while the emission is proportional to the square of the density. If one makes
the simplifying assumption that the same gas is responsible for both absorption
and emission, their ratio can be used to determine the electron density in the
gas \citep{ss}.

The best case in our sample for comparing emission and absorption measurements
would be sight line P10411 ($l=229\fdg3$), as it lies only 3\farcm5 from
WD~0455-282, which was included in the \ion{O}{6} absorption survey of
\citet{sav06}. However, the measured velocity of the \ion{O}{6} emission in the
P10411 spectrum is inconsistent with the interstellar component of the
\ion{O}{6} absorption toward WD~0455-282. If the observed \ion{O}{6} emission is
indeed of circumstellar origin as mentioned in \S\ \ref{sam}, it could be part
of a HIPN.

Finding suitable pairs of emission/absorption measurements is difficult due to
the clumpy nature of the \ion{O}{6}-bearing gas and the resulting strong
variations in absorption and emission on scales of less than one degree.
Fortunately, the extended \ion{O}{6} emission survey by Dixon, Sankrit, \& Otte
(2006) yielded two such pairs. With a growing archive of observations, it will
be useful to search for more emission/absorption pairs in the future.

\subsection{\label{hac}Comparison between \ion{O}{6} and H$\alpha$ Emission}

\begin{figure}
\epsscale{1.0}
\plotone{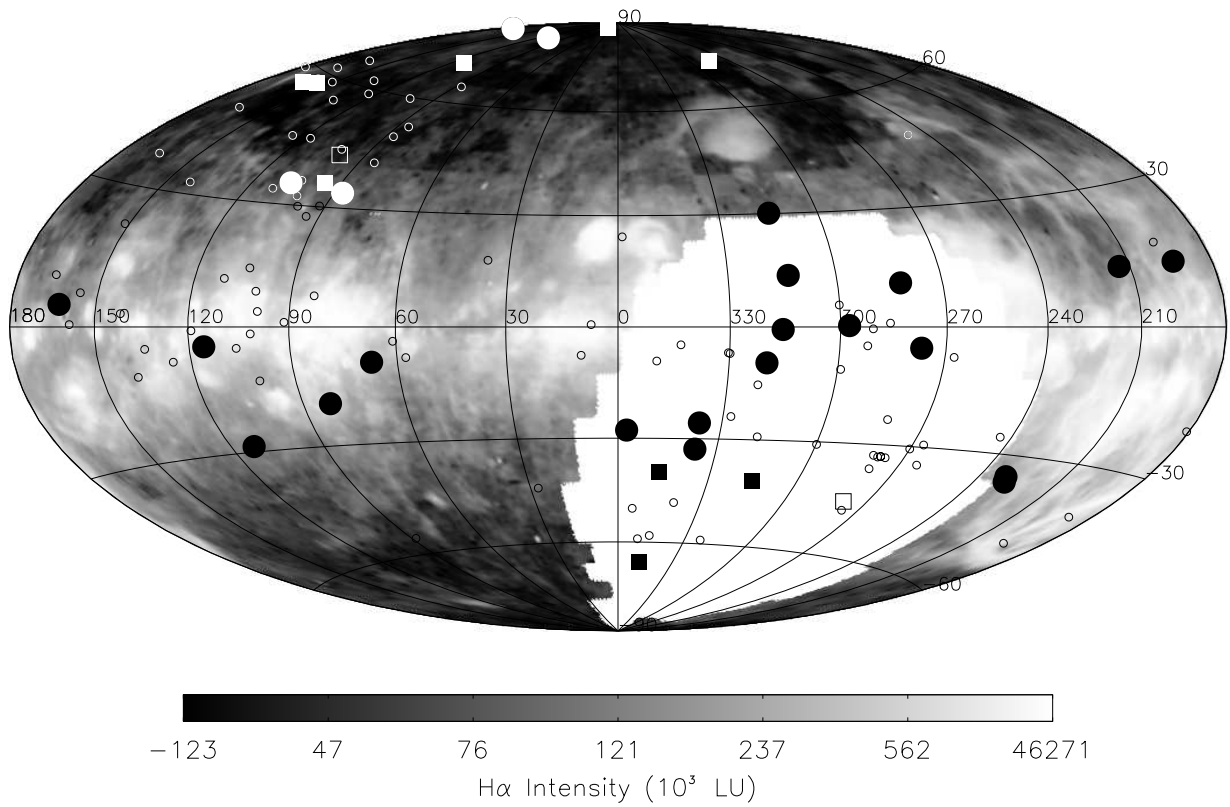}
\caption{\label{wham} {\em FUSE} \ion{O}{6} emission survey sight lines overlaid
on a WHAM map. Symbols and projection are the same as in Figure \ref{pos}.}
\end{figure}

Figure \ref{wham} shows a map of H$\alpha$ emission produced from the Wisconsin
H-Alpha Mapper (WHAM) survey \citep{haf} overlaid with the positions of our
survey sight lines and previously published sight lines. We used the Southern
H-Alpha Sky Survey Atlas \citep[SHASSA;][]{gau} to look for H$\alpha$ features
in the area not covered by the WHAM survey. SHASSA has a higher resolution than
the WHAM survey, but less sensitivity, so faint structures in the ionized gas
cannot be indentified with SHASSA. According to the WHAM and SHASSA maps,
\ion{O}{6} emission is detected in all environments, i.e., toward \ion{H}{2}
regions, filaments, and bubbles as well as faint, featureless ionized gas.
Tables \ref{det}--\ref{pub} include the H$\alpha$ intensities integrated over
the velocity range of $-80$\,km\,s$^{-1}<v<+80$\,km\,s$^{-1}$ measured by WHAM.

\begin{figure}
\plotone{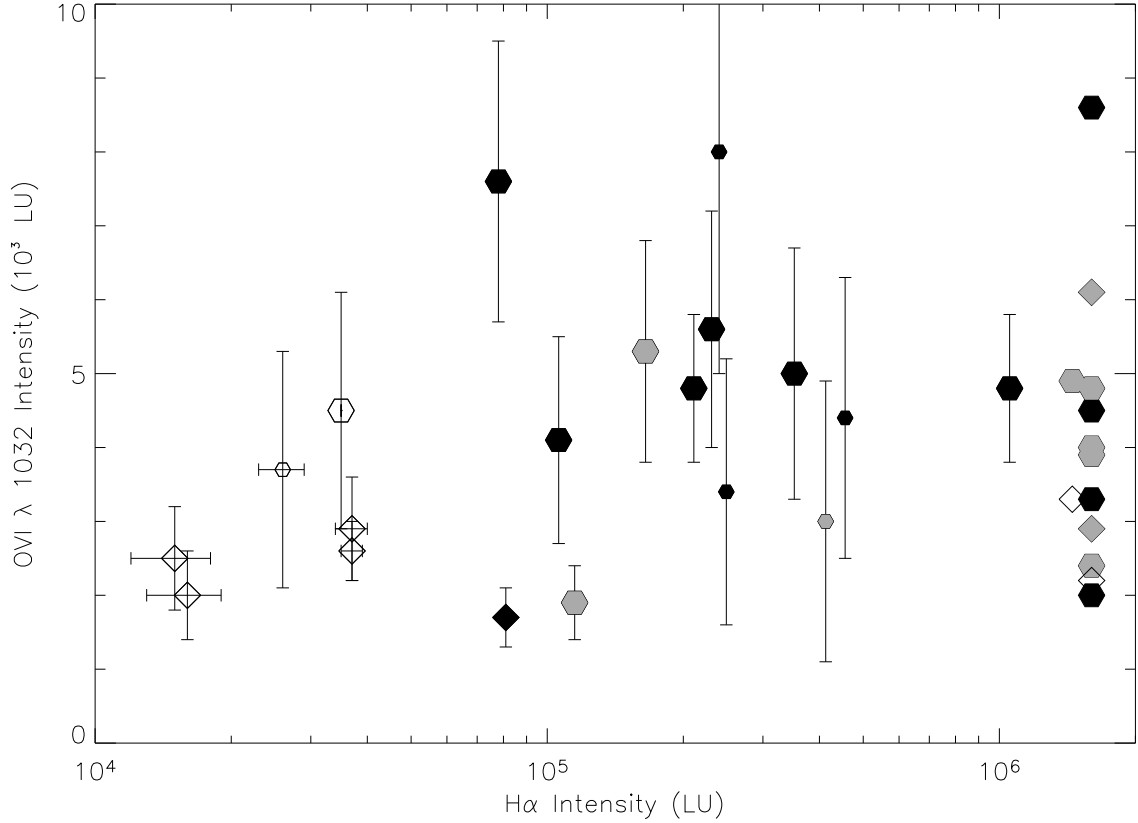}
\caption{\label{ha} Comparison between \ion{O}{6} and H$\alpha$ intensities.
Only detection sight lines are plotted. Hexagons mark the survey sight lines,
diamonds represent published sight lines from the literature. Black symbols show
detections toward structures in H$\alpha$ emission (e.g., filaments, \ion{H}{2}
regions) in the disk, gray and open symbols represent sight lines probing
diffuse gas in the disk and halo, respectively. The smaller symbols mark sight
lines with $I/\sigma_{\rm I}<2.7$. Sight lines not included in the WHAM survey
are plotted at the right side of the graph without error bars for completeness.
The \ion{O}{6} intensities for these sight lines are the measured values, but
their H$\alpha$ intensity has no meaning.}
\end{figure}

\citet{sav} found that \ion{O}{6} absorption and H$\alpha$ emission are poorly
correlated. In Fig.\ \ref{ha}, we compare the measured \ion{O}{6} $\lambda1032$
intensities with the H$\alpha$ intensities for all sight lines observable by
WHAM. The two instruments used different size apertures (a square of
900\,arcsec$^2$=0.25\,arcmin$^2$ for {\em FUSE}, a one-degree diameter or
2827\,arcmin$^2$ beam for WHAM); however, on a global (Galactic) scale, a
comparison between these two data sets is still useful. We divided the sight
lines into three groups represented by the differently shaded symbols in Fig.\
\ref{ha}: black symbols indicate sight lines that probe H$\alpha$ structures 
like filaments or bubbles in the Galactic disk (i.e., $|b|<45\degr$), gray and
open symbols represent sight lines through diffuse gas in the disk and halo,
respectively. Our survey does not include sight lines toward H$\alpha$
structures in the halo (i.e., at $|b|>45\degr$). (The H$\alpha$ morphology along
our 23 detection sight lines is described in Table \ref{locs}.)

Lower latitude sight lines most likely probe gas in the disk, since absorption
at low $b$ should prevent longer path lengths and thus the probing of halo gas.
In addition, H$\alpha$ features like filaments or bubbles can be clues that
different processes are at work in those regions than in diffuse, featureless
gas. We therefore calculated the average \ion{O}{6} and H$\alpha$ intensities
for the three groups of sight lines defined above. Combining the measurements
from Tables \ref{det} and \ref{pub} (that are covered by the WHAM survey) yields
an average \ion{O}{6} intensity of gas with H$\alpha$ structures in the disk of
$\bar{I}=4800\pm300$\,LU (with a SD of 1800\,LU around the mean). For diffuse
gas, $\bar{I}=3600\pm1200$\,LU (SD: 2400\,LU) in the disk and $2900\pm200$\,LU
(SD: 1000\,LU) in the halo. Again, no H$\alpha$ features are observed along halo
sight lines. The average H$\alpha$ intensities for these categories are
$3.0\pm0.5\times10^5$\,LU (SD: $3.5\times10^5$\,LU), $1.4\pm0.2\times10^5$\,LU
(SD: $3.5\times10^4$\,LU), and $2.8\pm0.2\times10^4$\,LU (SD: $1.1\times
10^4$\,LU), respectively. Both average \ion{O}{6} and H$\alpha$ intensities are
lowest for the halo sight lines and highest for the disk gas that exhibits
H$\alpha$ features. The halo sight lines clearly occupy a distinct area in Fig.
\ref{ha}. Including the sight lines for which no WHAM data exist yields an
average \ion{O}{6} intensity toward H$\alpha$ structures in the disk of
$4700\pm600$\,LU (SD: 2100\,LU) and, for diffuse gas, $\bar{I}=4000\pm500$\,LU
(SD: 1400\,LU) in the disk and $2900\pm300$\,LU (SD: 800\,LU) in the halo.

The {\em FUSE} and WHAM instruments possess different velocity resolutions
(106\,km\,s$^{-1}$ for the {\em FUSE} LWRS aperture, 2\,km\,s$^{-1}$ for WHAM).
Most of the H$\alpha$ emission falls within one resolution element of the {\em
FUSE}\/ observation. We therefore cannot distinguish individual components in
the \ion{O}{6} line profile; only the bulk emission can be compared in the
velocity range covered by WHAM ($|v_{\rm LSR}|\approx100$\,km\,s$^{-1}$). Of the
14 sight lines for which WHAM data exist, the \ion{O}{6} emission line fully
overlaps the H$\alpha$ emission in 11 of them. Eight of these show differences
between the \ion{O}{6} and H$\alpha$ emission line centers of less than
25\,km\,s$^{-1}$. The remaining three sight lines exhibit differences between 35
and 60\,km\,s$^{-1}$. Thus, the \ion{O}{6} and H$\alpha$ emission features
appear to have similar velocities along 2/3 of the sight lines (where data are
available), but {\em FUSE}\/ lacks the velocity resolution to conclude how much
of the H$\alpha$ emission originates in the same cloud as the \ion{O}{6}
emission.

Several comparisons between model predictions and observed column densities of
high-ionization species like \ion{O}{6} and \ion{C}{4} have been published
\citep[e.g.,][]{sem92, sem93, spi, indb}. Very few models, unfortunately, make
predictions about \ion{O}{6} emission. Under the simplifying assumption that all
of the \ion{O}{6} and H$\alpha$ emission along each sight line originates in the
same gas cloud, we compared their intensities with those predicted by models of
turbulent mixing layers (TML) and cooling Galactic Fountain gas. Slavin, Shull,
\& Begelman (1993) give \ion{O}{6} and H$\alpha$ intensities for TML models of
different abundances, temperatures, and velocities. The predicted H$\alpha$
intensities vary between 35 and 800\,LU, the \ion{O}{6} intensities go up to
about 12\,LU. This means that the predicted TML intensities are several orders
of magnitude smaller than the observed ones. Benjamin \& Shapiro (1996, private
communication) predict H$\alpha$ intensities for clouds of various sizes of
cooling Galactic Fountain gas between 2 and $2.5\times10^4$\,LU, which is the
intensity range covered by our halo sight lines. The corresponding \ion{O}{6}
intensities, however, are between 2 and $3\times10^4$\,LU, i.e. about ten times
higher than the observed intensities of the halo sight lines. Correcting the
observations for the low attenuation toward the halo cannot explain this
discrepancy. A single model therefore cannot explain the observed \ion{O}{6}
intensities. Absorption line studies \citep[e.g.,][]{spi, indb} also concluded
that mixed models are necessary to explain the observed \ion{O}{6} column
densities.
 
\subsection{\label{sxrc}Comparison between \ion{O}{6} and Soft X-Ray Emission}

\begin{figure}
\plotone{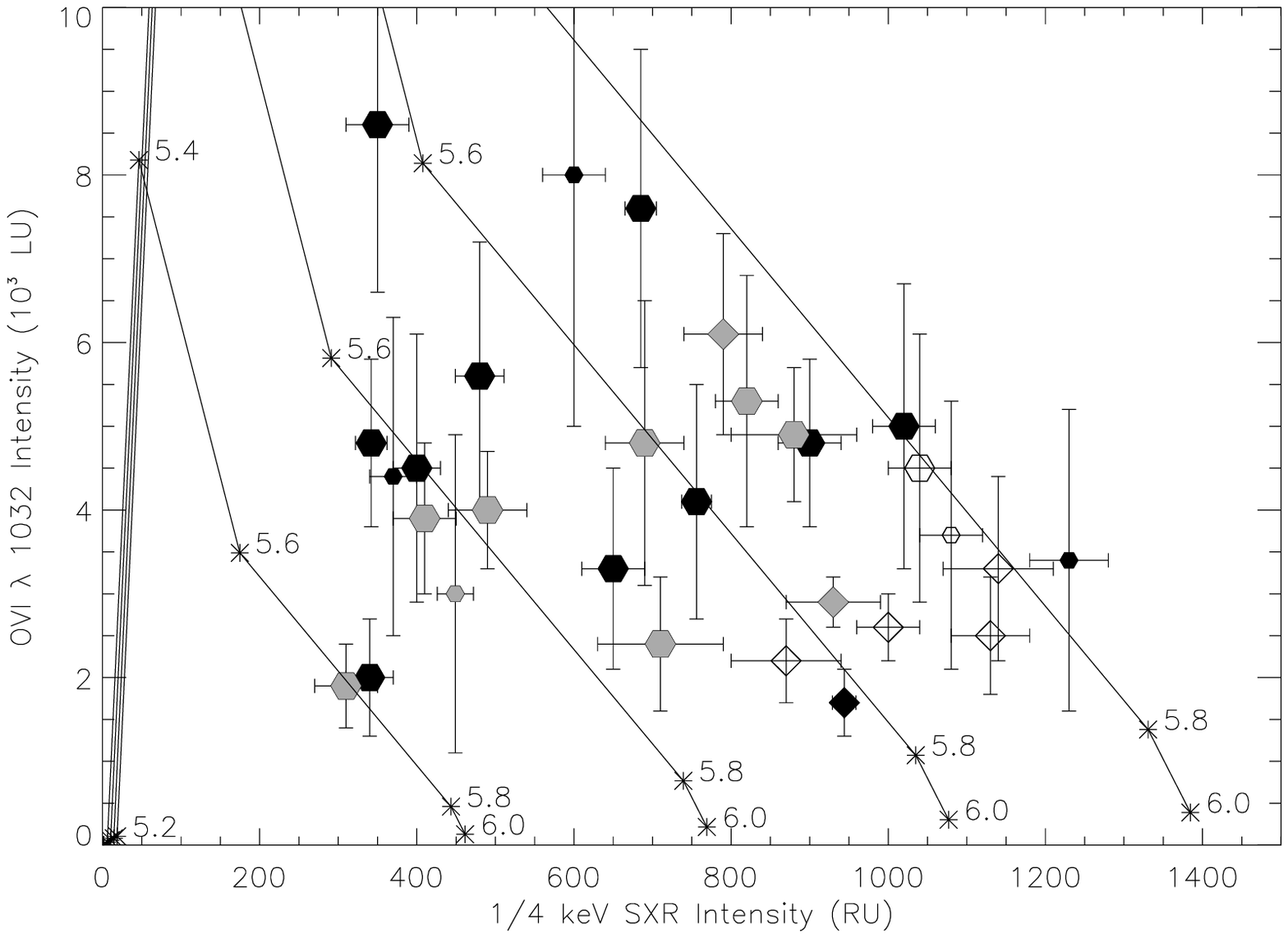}
\caption{\label{sxr} Comparison between \ion{O}{6} and SXR emission. Symbols are
the same as in Figure \ref{ha}. Only detection sight lines are shown. Sight
lines SL1 and SL2 (Table \ref{pub}) are not included (see text). The lines
combine intensities (asterisks) derived from the computed thermal spectra of
\citet{lan} for emission measures 0.015, 0.025, 0.035, and 0.045\,cm$^{-6}$\,pc
(increasing toward the upper right). The model intensities are labelled with
their corrsponding temperatures $\log(T)$.}
\end{figure}

One possible origin of the \ion{O}{6} emission is gas cooling from temperatures
of $10^6$\,K or more. Such hot gas is observable in the SXR regime. In Fig.\
\ref{sxr}, we compare the measured \ion{O}{6} $\lambda$1032 intensities with the
{\em ROSAT} 1/4\,keV SXR emission. Two of the \ion{O}{6} detections of Table
\ref{pub} (SL1 and SL2) are not included in the plot, as their unusually strong
SXR emission probably includes emission from the galaxy clusters located in
these directions.

If we divide the SXR data from Tables \ref{det} and \ref{pub} (excluding SL1 and
SL2) into three bins with an equal number of sight lines, the resulting bins
cover the SXR intensity ranges $<599$, 600--899, and $>900$\,RU ({\em ROSAT}
unit; 1\,RU = $10^{-6}$~counts~s$^{-1}$\,arcmin$^{-2}$). The mean \ion{O}{6}
intensities for each bin are $4400\pm800$\,LU (SD: 2100\,LU), $4500\pm600$\,LU
(SD: 1700\,LU), and $3400\pm400$\,LU (SD: 1200\,LU). Although the scatter in the
data is large, there appears to be a decrease in \ion{O}{6} intensity in the
regions of strongest SXR emission. If we divide the sight lines again into the
three categories mentioned in \S\ \ref{hac} (disk gas with and without H$\alpha$
features, diffuse halo gas) and exclude SL1 and SL2, the SXR intensities are
$620\pm80, 670\pm70$, and $1040\pm50$\,RU with SDs of 260, 220, and 110\,RU,
respectively. The average \ion{O}{6} intensity of this sample of halo sight
lines is $3000\pm400$\,LU with a SD of 900\,LU, i.e., only slightly different
from the value mentioned in \S\ \ref{hac}. Again, the areas of strongest SXR
emission exhibit the lowest \ion{O}{6} intensities.

\citet{lan} computed spectra emitted by hot ($10^4-10^8$\,K), optically thin gas
for the X-ray and ultraviolet wavelength regime assuming collisional ionization
equilibrium. We derived SXR and \ion{O}{6} intensities from their tables for
various temperatures and emission measures \citep[using a conversion factor of
$5.8\times10^{-12}$\,ergs\,cm$^{-2}$\,count$^{-1}$ for the {\em ROSAT}\/ SXR
counts;][]{fle}. The lines in Fig.\ \ref{sxr} show the resulting
\ion{O}{6} and SXR intensities for emission measures of 0.015, 0.025, 0.035, and
0.045\,cm$^{-6}$\,pc for various temperature values $\log(T)$. Observations
along the sight lines probing diffuse gas in the disk lie mostly in the central
part of the parameter space in Fig.\ \ref{sxr}, while the observations toward
the halo occupy the high SXR/low \ion{O}{6} intensity region. A comparison
between the model by \citet{lan} and the data is problematic mainly for three
reasons: 1) The model assumption of collisional ionization equilibrium does not
apply to cooling and possibly condensing Galactic Fountain gas. 2) The observed
data in the disk are affected by attenuation, a correction of which may change
the distribution of the data points in Fig.\ \ref{sxr}. While attenuation
certainly has an effect on the lower latitude sight lines (see \S\ \ref{tau}
below), it cannot explain the drop in \ion{O}{6} emission for the halo sight
lines. 3) The SXR intensities contain a contribution from the Local Bubble,
whereas the amount of \ion{O}{6} emission from the Local Bubble is still being
debated. Despite these drawbacks in our comparison, however, it is comforting to
see that the predicted and observed intensities are of the same order of
magnitude -- unlike the models in the \ion{O}{6}--H$\alpha$ comparison in \S\
\ref{hac} above.

\subsection{\label{rot}Galactic Rotation}

\begin{figure}
\plotone{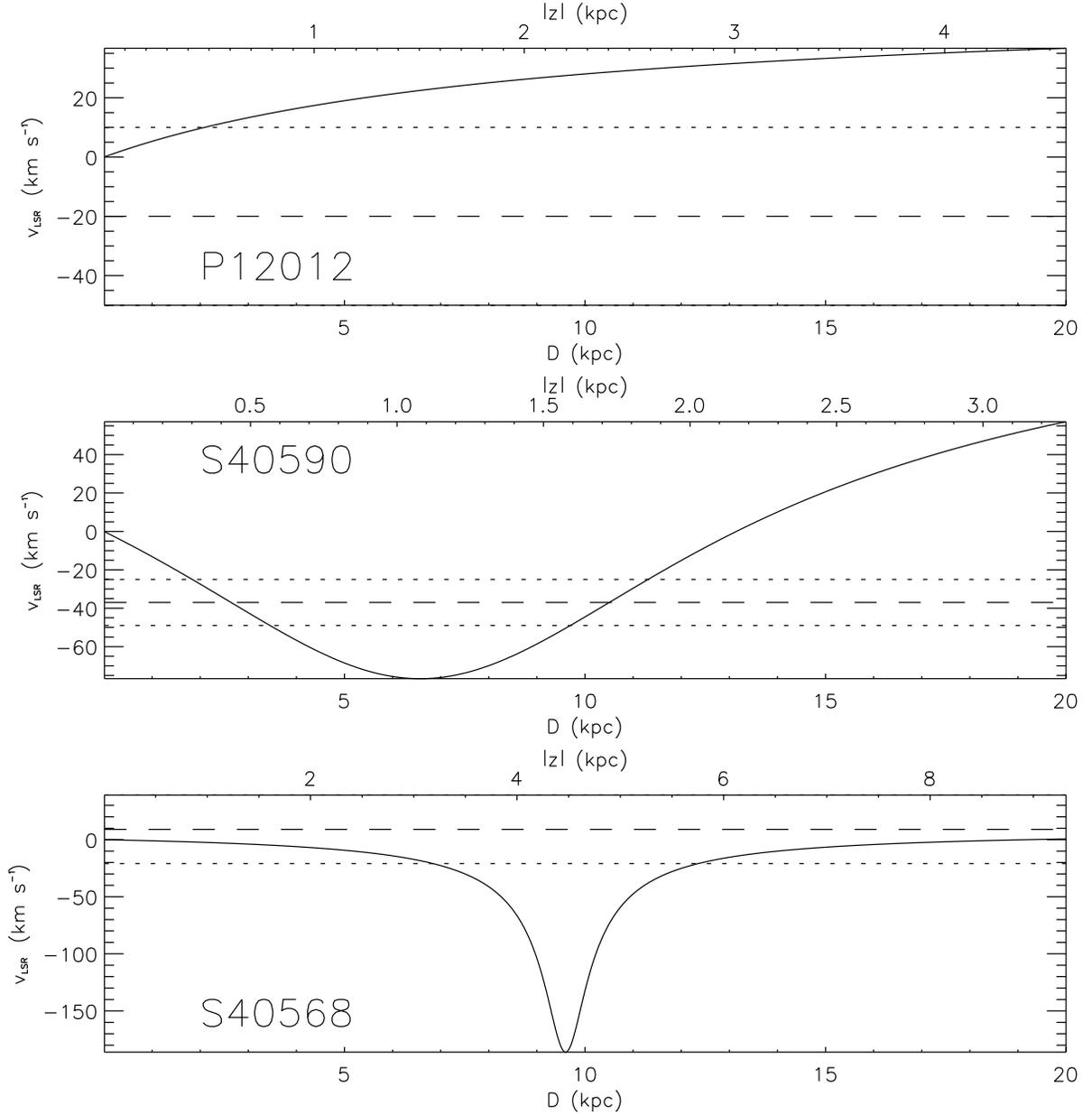}
\caption{\label{velp} Examples of model velocity profiles. For three of the 23
detection sight lines, the expected velocities of a corotating halo model are
plotted (solid lines). The measured velocity (dashed line) and the full
1$\sigma$ velocity range (dotted lines) are also shown. (Missing lines are
overlapping with an x-axis.) The bottom x-axis shows the distance $D$ along the
sight line, the top x-axis gives the height $|z|$ from the plane. The sample
profiles show almost no match between observation and model (top panel), two
solutions at different distances from the sun (middle panel), and a match along
almost the entire line of sight up to the end of the disk (bottom panel).}
\end{figure}

Table \ref{det} includes the local-standard-of-rest velocities $v_{\rm LSR}$ for
all \ion{O}{6} $\lambda$1032 detections. To compare these velocities with
Galactic rotation, we adopted a simple rotation model, a differentially
corotating halo with a constant velocity of $v=220$\,km\,s$^{-1}$. We derived
the distance $D$ along the line of sight to the \ion{O}{6} emitting gas and its
height $z$ above the Galactic plane from the match between the model and the
measured velocity. The values for $z$ and $D$ derived from our model are listed
in Table \ref{locs}. The quoted uncertainties in $z$ and $D$ are based on the
overlap between the 1$\sigma$ uncertainties of the measured velocities and the
velocities allowed by the corotation model. If a match between the measured
velocity and the model exists only within the uncertainty of the velocity, the
values for $z$ and $D$ are quoted as ranges. For sight lines with
$0\degr<l<90\degr$ or $270\degr<l<360\degr$, two solutions might exist for a
part of or the full 1$\sigma$ range, as the velocity contours of the corotation
model loop around inside the solar orbit around the Galactic center. There are
a handful of sight lines for which no solution exists within the observed
1$\sigma$ range. Two sight lines yield solutions up to the end of the Galactic
disk/halo (represented by question marks in Table \ref{locs}, since the
``exact'' end of the disk/halo is not known). Figure \ref{velp} shows the model
velocity profiles, the measured velocities, and their uncertainty ranges for
three survey detection sight lines as an example. The model velocities were
always calculated for $D=0-20$\,kpc without consideration of the direction of
the sight line.

The measured velocities of seven of the 23 sight lines are inconsistent with the
corotating halo model (e.g., measured $v_{\rm LSR}>0$ in $v_{\rm halo}<0$ region
or vice versa). The other sight lines (about 2/3 of the detection sight lines)
yield solutions of various likelihoods. These depend on the probability of the
observed velocity and the derived values for $|z|$ and $D$ with regard to the
expected extinction along and the direction of the sight line. For sight lines
with two solutions, for example, the first listed solution is usually more
likely due to the higher extinction along the longer path length and the larger
(and sometimes unreasonable) values for $|z|$ or $D$ of the second solution. The
comment in Table \ref{locs} takes these probabilities into account and describes
what we believe is the likely range of the corotating, emitting gas. We define
``local'' as $D\leq500$\,pc, ``near'' as $D\leq5$\,kpc, and (thick) disk gas as
$|z|\leq2.3$\,kpc. All sight lines yielding solutions for the corotation model
place the emitting gas in the near disk, in a few cases even in the local
environment. None of the sight lines yields only a halo solution (with halo now
defined as $|z|>2.3$\,kpc) without a near disk solution.

The measured velocities vary particularly strongly toward the northern Galactic
pole ($b>60\degr$). Combining the detections of Tables \ref{det} and \ref{pub},
we find five sight lines that fall into that region: S40581 ($l=110\fdg3$,
$I/\sigma_{\rm I}=2.3$), S40508 ($l=175\fdg0$), SL1, SL2, and SL4. Positive as
well as negative \ion{O}{6} $\lambda$1032 velocities are measured in this
direction, suggesting that turbulence, outflow, and infall play an important
role in the distribution of the \ion{O}{6}-emitting gas. One caveat, however, is
the low spectral resolution of our 106\,km\,s$^{-1}$ wide aperture, which would
be unable to resolve the emission from multiple clouds along the line of sight
(due to the strong binning of the data). If multiple clouds were present, the
derived values for $v_{\rm LSR}$ would represent an average for each sight line.

\subsection{\label{hvc}High-Velocity Clouds}

A large number of high-velocity ($|v|>100$\,km\,s$^{-1}$) components was
detected in the \ion{O}{6} absorption survey of \citet{sem}. The majority of
these components was observed toward high-velocity \ion{H}{1} 21\,cm emission.
Using the maps of high-velocity \ion{H}{1} gas in \citet[][their Fig. 16]{wak03}
and \citet[their Fig.\ 11]{sem}, we identified the sight lines in our \ion{O}{6}
emission survey that intersect \ion{H}{1} HVCs. Almost half of the sight lines
in our survey (52 out of 112) pass regions of high-velocity \ion{H}{1} gas, yet
of the 23 sight lines showing \ion{O}{6} emission, only five intersect
\ion{H}{1} HVCs (all with $I/\sigma_{\rm I}\geq2.7$). In none of these cases do
the measured \ion{O}{6} and \ion{H}{1} velocities agree ($|v_{\rm OVI}-v_{\rm
HI}|>50$\,km\,s$^{-1}$); however, we have found one region of the sky that
exhibits \ion{O}{6} emission and absorption at similar velocities. Sight line
S40521 ($l=81\fdg9$, $I/\sigma_{\rm I}=2.3$) shows a weak emission line at
$v_{\rm LSR}=-320\pm30$\,km\,s$^{-1}$. Three nearby absorption sight lines,
Mrk1513, Mrk304, and UGC12163, exhibit high-velocity \ion{O}{6} components at
$v_{\rm LSR}=-293, -304,$ and $-274$\,km\,s$^{-1}$, respectively. \citet{sem}
identify this region as an extension of the Magellanic Stream. Except for this
region of the Magellanic Stream, our survey does not contain any sight lines
near HVCs originally detected in high-ionization species such as \ion{C}{4} or
\ion{O}{6} and studied by \citet{sem95,sem} and Collins, Shull, \& Giroux
(2005). Detecting \ion{O}{6} emission from HVCs appears more difficult than
detecting diffuse \ion{O}{6} emission -- given the fact that none of the 52
sight lines that pass regions of \ion{H}{1} HVCs exhibits \ion{O}{6} emission
from the HVC. In their extended survey, \citet{dix06} found two (of a total of
183) sight lines with high-velocity \ion{O}{6} emission matching the
high-velocity \ion{H}{1} emission.

\subsection{\label{tau}Variation in Extinction and \ion{O}{6} Intensity with
Galactic Latitude}

\begin{figure}
\plotone{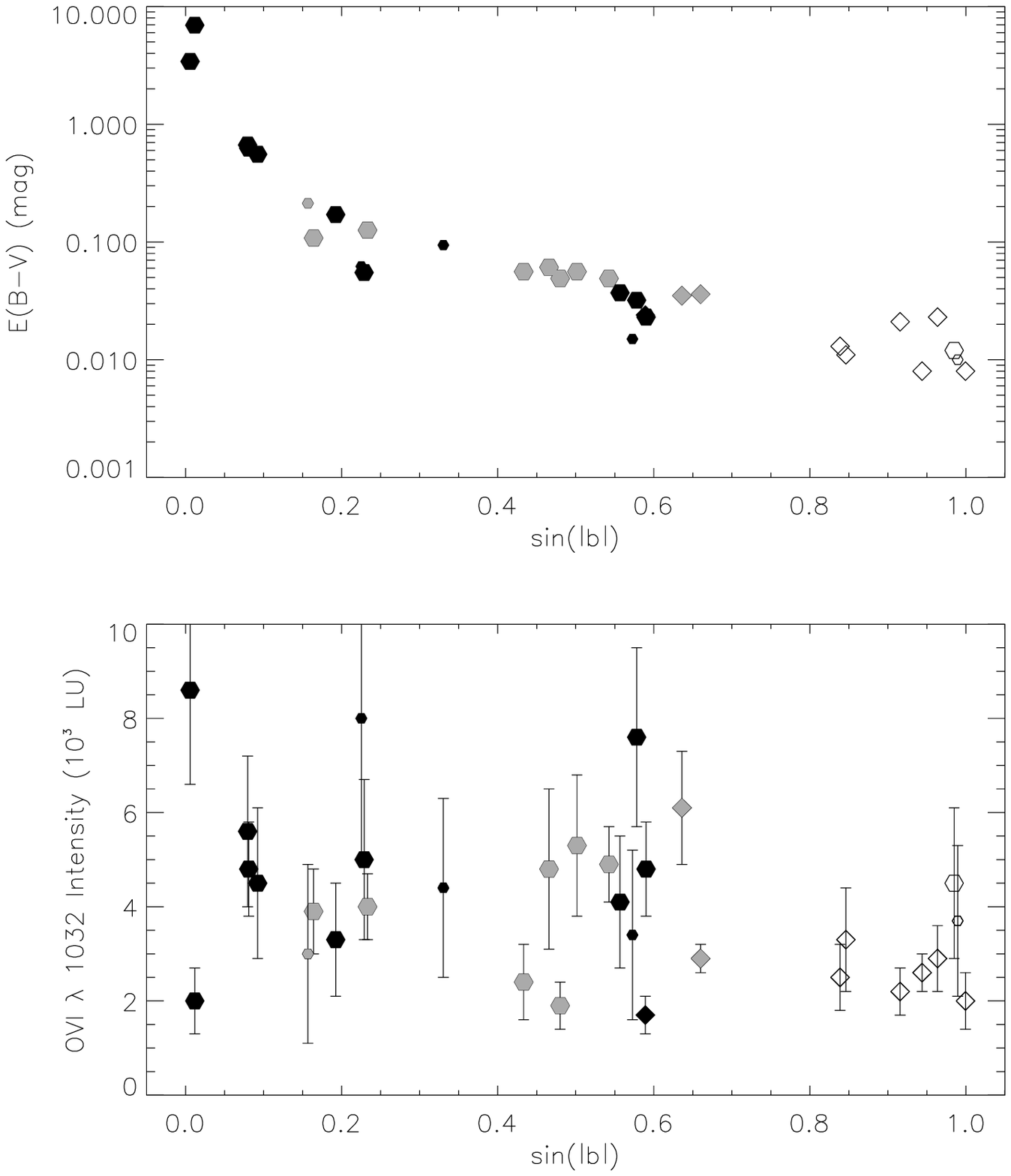}
\caption{\label{sinb} {\em Upper panel:} Variation of $E(\bv)$ with Galactic
latitude. {\em Lower panel:} Variation of \ion{O}{6} intensity with Galactic
latitude. Symbols are the same as in Figure \ref{ha}. Only detection sight lines
are shown.}
\end{figure}

When analyzing emission measurements in the ultraviolet, it is important to
consider the effects of interstellar extinction. A plot of the color excess
$E(B-V)$ versus $\sin(|b|)$ (where $b$ is the Galactic latitude) for sight lines
showing \ion{O}{6} emission is presented in the upper panel of Fig.\ \ref{sinb}.
The sharp decrease in dust extinction with Galactic latitude is obvious. The
high color excess at low latitudes suggests strongly that the \ion{O}{6}
emission is local (that is, closer than the obscuring dust).

The observed \ion{O}{6} $\lambda$1032 intensities are plotted against $\sin|b|$
in the lower panel of Fig.\ \ref{sinb}. For the nine high signal-to-noise sight
lines with $\sin|b|<0.4$, the average intensity is $4600\pm600$\,LU (SD:
1800\,LU). For the 11 sight lines with $0.4<\sin|b|<0.7$, the average intensity
is $4200\pm600$\,LU (SD: 1800\,LU). For the seven remaining sight lines, with
$\sin|b|>0.7$, the average intensity is $2900\pm300$\,LU (SD: 800\,LU). We see
that, at high latitudes, the intensities are lower and the dispersion decreases.
Based on these plots, we divide our sample into low-latitude, mid-latitude, and
high-latitude sight lines. While both extinction and observed \ion{O}{6}
intensity fall with increasing Galactic latitude for our complete sample, within
each group we see no correlation of the observed \ion{O}{6} intensity with
reddening.

The relatively low dispersion in the intensities of our high-latitude sight
lines suggests that they sample gas with a uniform set of properties. The
\ion{O}{6} emission observed at these latitudes is thus likely to be from hot
gas in the thick disk/halo. The low-latitude sight lines are significantly
brighter than at higher latitudes, even though their extinction is higher;
hence, the low-latitude emission must come from nearby regions, probably
interfaces in the thin disk, as suggested by the absorption data. The average
intensity for the mid-latitude sight lines lies between the averages for the
high and low-latitude sight lines. This is simply explained if at these
latitudes the sight lines pass through both nearby emission interfaces and more
distant \ion{O}{6} in the thick disk. The dispersion then reflects the
clumpiness of the hot gas in the Galaxy.

\section{\label{conc}CONCLUSIONS}

We have examined 112 sight lines from the {\em FUSE} archive that probe diffuse,
hot gas in the Milky Way and found 23 that exhibit \ion{O}{6} $\lambda$1032
emission. Sight lines crossing morphological features like H$\alpha$ filaments
or bubbles in the disk exhibit the strongest \ion{O}{6} emission. Halo sight
lines show the lowest \ion{O}{6} and H$\alpha$ intensities, the strongest SXR
emission, and little or no H$\alpha$ structure. This suggests that different
mechanisms are at work in areas that show morphological features as opposed to
purely diffuse gas. Higher \ion{O}{6} intensities toward H$\alpha$ filaments and
bubbles indicate that the \ion{O}{6} emission originates in interfaces or
turbulent mixing layers. Neither TML nor Galactic Fountain models can reproduce
the observed intensities. Data derived from computed thermal X-ray/UV spectra,
however, are consistent with our observations for a range of emission measures
and temperatures. About 2/3 of our detection sight lines probe clouds whose
velocities are consistent with a corotating halo, placing the emitting gas most
likely within a few kpc in the thick disk, if not locally within only a few
100\,pc from the sun. Sight lines toward the northern Galactic pole
($b>60\degr$) show velocities that differ from the corotation model and indicate
turbulences, outflows, or infall of gas. None of our sight lines shows
\ion{O}{6} emission originating in \ion{H}{1} HVCs. Only one sight line
possesses \ion{O}{6} emission coinciding with the detection of high-velocity
\ion{O}{6} absorption toward the extension of the Magellanic Stream. Combined
with values from the literature, the \ion{O}{6} intensities from our survey
exhibit a variation with latitude that is consistent with the picture derived
from recent \ion{O}{6} absorption measurements: high-latitude sight lines probe
\ion{O}{6}-emitting gas in a clumpy, thick disk, while low-latitude sight lines
sample mostly mixing layers and interfaces in the thin disk of the Galaxy. By
combining \ion{O}{6} intensities with column densities derived from
absorption-line surveys, we can constrain the local density and physical scale
of the emitting gas. The challenge is finding suitable emission/absorption
measurements due to the clumpy nature of the \ion{O}{6}-bearing gas. Theoretical
models of this phase of the interstellar medium in the Galaxy must reproduce
both the large-scale distribution of \ion{O}{6} emission and the small-scale
microphysics derived from \ion{O}{6} emission and absorption surveys. No single
model can explain the observed emission and absorption data at this point.

\acknowledgements

The authors are grateful to R. Sankrit for his useful suggestions. The authors
also thank L. M. Haffner for his technical support on the WHAM data and R. J.
Reynolds for his useful comments. The authors are grateful to B. P. Wakker for
his help regarding the high-velocity cloud survey. This research is supported by
NASA contract NAS5-32985 to the Johns Hopkins University. B. Otte also
acknowledges support by grant NAG5-10765. The data presented in this paper were
obtained from the Multimission Archive at the Space Telescope Science Institute
(MAST). STScI is operated by the Association of Universities for Research in
Astronomy, Inc., under NASA contract NAS5-26555. Support for MAST for non-HST
data is provided by the NASA Office of Space Science via grant NAG5-7584 and by
other grants and contracts. The Wisconsin H-Alpha Mapper is funded by the
National Science Foundation. This work made use of the NASA Astrophysics Data
System (ADS) and the NASA Extragalactic Database (NED).


\clearpage

\begin{deluxetable}{ll}
\tablewidth{0pt}
\tablecaption{\label{obsid} OBSERVATIONS}
\tablehead{\colhead{Target ID} & \colhead{Observation IDs}}
\startdata
A01002   & A0100201 \\
A03406   & A0340601 \\
A03407   & A0340701 \\
A04604   & A0460404 \\
A04802   & A0480202 \\
A04905   & A0490501, A0490502 \\
A05101   & A0510102 \\
A10001   & A1000101 \\
A11101   & A1110101 \\
A11102   & A1110201 \\
B01805   & B0180501 \\
B04604   & B0460401, B0460501 \\
B06801   & B0680101 \\
B09401   & B0940101 \\
C02201   & C0220101, C0220102 \\
C02301   & C0230101 \\
C03701   & C0370101 \\
C03702   & C0370201, C0370202, C0370203, C0370204, C0370205 \\
C05501   & C0550101 \\
C07602   & C0760201, C0760301 \\
C07604   & C0760401 \\
M10310   & M1031003, M1031004, M1031006, M1031007 \\
M10703   & M1070301, M1070302, M1070304, M1070305, M1070307, M1070308, \\
         & M1070310, M1070311, M1070313, M1070314, M1070319 \\
P10407   & P1040701 \\
P10409   & P1040901 \\
P10411   & P1041101, P1041102, P1041103 \\
P10413   & P1041303 \\
P10414   & P1041403 \\
P10418   & P1041801 \\
P10425   & P1042501, P1042601 \\
P12012   & P1201112, P1201213 \\
P20406   & P2040601 \\
P20410   & P2041002, S4053601 \\
P20411   & P2041102, P2041103, P2041104 \\
P20419   & P2041901 \\
P20421   & P2042101 \\
P20422   & P2042201, P2042202, P2042203, S4055602, S4055603, S4055607, \\
         & S4055608, S4055609 \\
P20423   & P2042301 \\
P20517   & P2051701, P2051702, P2051703 \\
Q10803   & Q1080303 \\
Q11002   & Q1100201 \\
S30402   & S3040203, S3040204, S3040205, S3040206, S3040207, S3040208 \\
S40502   & S4050201 \\
S40504   & S4050401 \\
S40505   & P1042101, P1042105, S4050501, S4050502, S4050503 \\
S40506   & S4050601 \\
S40507   & S4050701 \\
S40508   & S4050801 \\
S40509   & S4050903 \\
S40510   & S4051001, S4051002 \\
S40512   & S4051201 \\
S40513   & S4051301 \\
S40514   & S4051401 \\
S40518   & S4051801, S4051802, S4051803, S4051804, S4051805 \\
S40521   & S4052101 \\
S40522   & S4052201 \\
S40523   & S4052301, S4052302 \\
S40524   & S4052402, S4052403, S4052404, S4052405, S4052406 \\
S40525   & S4052501 \\
S40526   & S4052601 \\
S40529   & S4052901, S4052902 \\
S40531   & S4053101 \\
S40532   & S4053201, S4053202 \\
S40533   & S4053301 \\
S40535   & S4053501 \\
S40537   & S4053701, S4053702 \\
S40540   & S4054001, S4054002, S4054003, S4054004, S4054005 \\
S40541   & S4054101 \\
S40543   & S4054301 \\
S40553   & S4055301 \\
S40555   & S4055501, S4055502 \\
S40557   & S4055701, S4055702, S4055703, S4055704, S4055705 \\
S40558   & S4055801 \\
S40560   & S4056001, S4056002 \\
S40562   & S4056201 \\
S40563   & S1010206, S4056301, S4056302, S4056303, S4056304 \\
S40564   & S4056401 \\
S40566   & S4056601, S4056602, S4056603 \\
S40568   & S4056801, S4056802 \\
S40569   & S4056901 \\
S40570   & S4057001 \\
S40572   & S4057201 \\
S40573   & S4057301, S4057302 \\
S40574   & S4057401 \\
S40576   & S4057601, S4057602, S4057603, S4057604 \\
S40579   & P1042901, P1042902, S4057901, S4057902, S4057903, S4057904 \\
S40580   & S4055001, S4058001 \\
S40581   & S4058101, S4058102 \\
S40582   & S4058201, S4058202 \\
S40587   & S4058701 \\
S40588   & S4058801 \\
S40589   & S4058901 \\
S40590   & S4059001 \\
S40591   & S4059101 \\
S40593   & S4059301 \\
S51303   & S5130301 \\
S51401   & S5140102, S5140103 \\
Z90702   & Z9070201 \\
Z90703   & Z9070301 \\
Z90708   & Z9070801 \\
Z90709   & Z9070901 \\
Z90711   & Z9071101 \\
Z90712   & Z9071201 \\
Z90714   & Z9071401 \\
Z90715   & Z9071501 \\
Z90719   & Z9071901 \\
Z90726   & Z9072601, Z9072602 \\
Z90727   & Z9072701 \\
Z90732   & Z9073202 \\
Z90735   & Z9073501 \\
Z90736   & Z9073601 \\
Z90737   & Z9073701, Z9073702 \\
\enddata
\end{deluxetable}

\clearpage

\begin{deluxetable}{lrrrcccccccccc}
\tabletypesize{\tiny}
\rotate
\tablewidth{0pt}
\tablecaption{\label{det} \ion{O}{6} $\lambda$1032 MEASUREMENTS}
\tablehead{\colhead{Target} & \colhead{$l$} & \colhead{$b$} &
\colhead{$t$\tablenotemark{a}} & \colhead{$I_{1032}$\tablenotemark{b}} &
\colhead{$\lambda_{\rm c}$\tablenotemark{c}} & \colhead{FWHM} & \colhead{} &
\colhead{$1-P_{\rm F}$} & \colhead{$v_{\rm LSR}$} & \colhead{} &
\colhead{SXR\tablenotemark{e}} & \colhead{$I_{{\rm H}\alpha}$\tablenotemark{f}}
& \colhead{} \\
\colhead{ID} & \colhead{(deg.)} & \colhead{(deg.)} & \colhead{(sec)} &
\colhead{(10$^3$\,LU)} & \colhead{(\AA)} & \colhead{(km\,s$^{-1}$)} &
\colhead{$\chi^2/\nu$} & \colhead{(\%)} & \colhead{(km\,s$^{-1}$)} &
\colhead{$E(B-V)$\tablenotemark{d}} & \colhead{(RU)} & \colhead{(10$^3$\,LU)} &
\colhead{HVC\tablenotemark{g}}}
\startdata
S40523   &  67.219 & $  -9.024$ &   9820 & $   3.0\pm  1.9$ & $1031.98 \pm0.13 $
 & $120  \pm150  $ & 0.94 & 56 & $  30  \pm40  $ & 0.213 & $  449\pm 23$ &
 $  413\pm 4$ & --- \\
S40521   &  81.875 & $ -19.293$ &  13539 & $   4.4\pm  1.9$ & $1030.78 \pm0.10 $
 & $140  \pm 90  $ & 1.08 & 66 & $-320  \pm30  $ & 0.094 & $  370\pm 30$ &
 $  456\pm 0$ & --- \\
S40560   &  87.172 & $+ 33.829$ &  19249 & $   4.1\pm  1.4$ & $1031.86 \pm0.07 $
 & $110  \pm 80  $ & 1.38 & 73 & $  -3  \pm19  $ & 0.037 & $  756\pm 19$ &
 $  106\pm 4$ & C   \\
Z90715   & 106.808 & $+ 35.314$ &  14576 & $   7.6\pm  1.9$ & $1031.85 \pm0.10 $
 & $200  \pm 30  $ & 1.61 & 76 & $ -10  \pm30  $ & 0.032 & $  685\pm 20$ &
 $   78\pm 3$ & C   \\
S40581   & 110.308 & $+ 81.653$ &  14330 & $   3.7\pm  1.6$ & $1031.70 \pm0.18 $
 & $180  \pm120  $ & 0.95 & 66 & $ -60  \pm50  $ & 0.010 & $ 1080\pm 40$ &
 $   26\pm 3$ & --- \\
P20411   & 112.482 & $ -28.688$ &  48281 & $   1.9\pm  0.5$ & $1031.82 \pm0.04 $
 & $  4  \pm 80  $ & 0.89 & 75 & $ -25  \pm12  $ & 0.049 & $  310\pm 40$ &
 $  115\pm 3$ & --- \\
S30402   & 115.438 & $  -4.644$ &  30491 & $   4.8\pm  1.0$ & $1032.62 \pm0.09 $
 & $201  \pm 19  $ & 0.82 & 84 & $ 210  \pm25  $ & 0.627 & $  342\pm 20$ &
 $ 1054\pm 0$ & H/G \\
P10413   & 162.590 & $+  4.566$ &  14644 & $   5.6\pm  1.6$ & $1031.84 \pm0.10 $
 & $180  \pm 60  $ & 1.00 & 77 & $ -30  \pm30  $ & 0.667 & $  480\pm 31$ &
 $  231\pm 0$ & --- \\
S40508   & 175.036 & $+ 80.024$ &   8290 & $   4.5\pm  1.6$ & $1032.08 \pm0.09 $
 & $110  \pm100  $ & 0.66 & 79 & $  50  \pm30  $ & 0.012 & $ 1040\pm 40$ &
 $   35\pm 0$ & --- \\
P12012   & 194.301 & $+ 13.230$ &   9680 & $   5.0\pm  1.7$ & $1031.89 \pm0.11 $
 & $200  \pm 50  $ & 0.94 & 64 & $ -20  \pm30  $ & 0.055 & $ 1020\pm 40$ &
 $  352\pm 3$ & --- \\
P10418   & 213.704 & $+ 13.019$ &   3350 & $   8  \pm  3  $ & $1031.75 \pm0.08 $
 & $ 80  \pm150  $ & 0.97 & 69 & $ -67  \pm24  $ & 0.062 & $  600\pm 40$ &
 $  240\pm 0$ & --- \\
P10411   & 229.296 & $ -36.167$ &  46120 & $   4.8\pm  1.0$ & $1031.96 \pm0.04 $
 & $100  \pm 50  $ & 0.97 & 98 & $ -10  \pm12  $ & 0.023 & $  900\pm 40$ &
 $  211\pm 3$ & --- \\
S40537   & 230.674 & $ -34.936$ &  21020 & $   3.4\pm  1.8$ & $1031.90 \pm0.07 $
 & $ 90  \pm170  $ & 1.09 & 73 & $ -26  \pm19  $ & 0.015 & $ 1230\pm 50$ &
 $  249\pm 4$ & --- \\
P20517   & 277.041 & $  -5.306$ &  35923 & $   4.5\pm  1.6$ & $1031.80 \pm0.11 $
 & $190  \pm120  $ & 1.33 & 73 & $ -50  \pm30  $ & 0.556 & $  400\pm 30$ &
 \nodata      & --- \\
S40506   & 282.142 & $+ 11.097$ &  12990 & $   3.3\pm  1.2$ & $1031.85 \pm0.09 $
 & $ 80  \pm 50  $ & 1.20 & 62 & $ -30  \pm30  $ & 0.171 & $  650\pm 40$ &
 \nodata      & WD? \\
S40591   & 297.557 & $+  0.337$ &   7131 & $   8.6\pm  2.0$ & $1031.97 \pm0.06 $
 & $ 80  \pm 60  $ & 1.21 & 69 & $   4  \pm18  $ & 3.417 & $  350\pm 40$ &
 \nodata      & --- \\
S40557   & 313.373 & $+ 13.487$ &  48836 & $   4.0\pm  0.7$ & $1031.74 \pm0.03 $
 & $ 77  \pm 23  $ & 0.96 & 98 & $ -57  \pm10  $ & 0.126 & $  490\pm 50$ &
 \nodata      & --- \\
A04604   & 314.861 & $+ 30.105$ &  18147 & $   5.3\pm  1.5$ & $1031.91 \pm0.07 $
 & $160  \pm 60  $ & 1.09 & 79 & $  -4  \pm21  $ & 0.056 & $  820\pm 40$ &
 $  165\pm 3$ & --- \\
P10425   & 315.732 & $  -0.679$ &  26568 & $   2.0\pm  0.7$ & $1031.92 \pm0.06 $
 & $ 10  \pm180  $ & 1.04 & 67 & $  -4  \pm16  $ & 6.950 & $  340\pm 30$ &
 \nodata      & WE? \\
S40590   & 319.714 & $  -9.448$ &  10914 & $   3.9\pm  0.9$ & $1031.81 \pm0.04 $
 & $ 10  \pm 40  $ & 0.88 & 78 & $ -37  \pm12  $ & 0.108 & $  410\pm 40$ &
 \nodata      & --- \\
C07602   & 336.501 & $ -25.677$ &  33042 & $   2.4\pm  0.8$ & $1031.89 \pm0.06 $
 & $ 50  \pm 50  $ & 1.44 & 69 & $ -10  \pm18  $ & 0.056 & $  710\pm 80$ &
 \nodata      & --- \\
P20422   & 336.583 & $ -32.859$ &  76990 & $   4.9\pm  0.8$ & $1031.87 \pm0.04 $
 & $140  \pm 30  $ & 1.13 & 99 & $ -17  \pm11  $ & 0.049 & $  880\pm 80$ &
 \nodata      & --- \\
S40568   & 357.478 & $ -27.762$ &  11912 & $   4.8\pm  1.7$ & $1031.94 \pm0.11 $
 & $180  \pm 80  $ & 1.16 & 64 & $   9  \pm30  $ & 0.061 & $  690\pm 50$ &
 \nodata      & --- \\
\enddata
\tablenotetext{a}{Exposure time is night only.}
\tablenotetext{b}{1\,LU = 1~photon~s$^{-1}$\,cm$^{-2}$\,sr$^{-1}$; for
\ion{O}{6} $\lambda$1032, 1\,LU =
$4.5\times10^{-22}$~erg~s$^{-1}$\,cm$^{-2}$\,arcsec$^{-2}$.}
\tablenotetext{c}{Wavelength is heliocentric.}
\tablenotetext{d}{Total reddening \citep{sfd}.}
\tablenotetext{e}{{\em ROSAT} $\onequarter$ keV emission with 1\,RU =
$10^{-6}$~counts~s$^{-1}$\,arcmin$^{-2}$ \citep{snow}.}
\tablenotetext{f}{H$\alpha$ intensity integrated over velocity range
$-80$\,km\,s$^{-1}<v<+80$\,km\,s$^{-1}$; an uncertainty of 0 means that the
intensity is the average of the neighboring pointings \citep{haf}; for
H$\alpha$, 1\,LU = $7.1\times10^{-23}$~erg~s$^{-1}$\,cm$^{-2}$\,arcsec$^{-2}$.}
\tablenotetext{g}{Based on \ion{H}{1} map in \citet{wak03}.}
\end{deluxetable}

\clearpage

\begin{deluxetable}{lrrrccccc}
\tabletypesize{\small}
\tablewidth{0pt}
\tablecaption{\label{upl} \ion{O}{6} $\lambda$1032 3$\sigma$ UPPER LIMITS}
\tablehead{\colhead{Target} & \colhead{$l$} & \colhead{$b$} &
\colhead{$t$\tablenotemark{a}} & \colhead{$I_{1032}$\tablenotemark{b}} &
\colhead{} & \colhead{SXR\tablenotemark{d}} & \colhead{$I_{{\rm
H}\alpha}$\tablenotemark{e}} & \colhead{} \\
\colhead{ID} & \colhead{(deg.)} & \colhead{(deg.)} & \colhead{(sec)} &
\colhead{(10$^3$\,LU)} & \colhead{$E(B-V)$\tablenotemark{c}} & \colhead{(RU)} &
\colhead{(10$^3$\,LU)} & \colhead{HVC\tablenotemark{f}}}
\startdata
S40512   &   7.153 & $+  0.616$ &   4280 &$   4          $ & 7.919 &
$  310\pm 40$ & $ 2583\pm 5$ & --- \\
Q10803   &   9.876 & $  -7.555$ &   3977 &$   3          $ & 0.331 &
$  270\pm 30$ & $  748\pm 4$ & --- \\
P20406   &  27.361 & $ -43.759$ &  18938 &$   1.9        $ & 0.047 &
$  720\pm 40$ & $   75\pm 0$ & --- \\
S40525   &  36.064 & $+ 17.621$ &   2840 &$   4          $ & 0.193 &
$  440\pm 30$ & $  225\pm 3$ & C/D \\
S40533   &  57.538 & $  -7.929$ &   5361 &$   3          $ & 0.301 &
$  540\pm 30$ & $  424\pm 4$ & --- \\
M10703   &  60.838 & $  -3.696$ &  59975 &$   2.4        $ & 1.415 &
$  560\pm 30$ & $  622\pm 0$ & --- \\
S40566   &  83.328 & $+  7.756$ &  20569 &$   2.0        $ & 0.459 &
$  488\pm 18$ & $ 1794\pm 5$ & --- \\
C03701   &  84.517 & $+ 42.189$ &  29877 &$   1.2        $ & 0.006 &
$ 1410\pm 40$ & $   38\pm 0$ & C   \\
S40555   &  85.357 & $+ 52.349$ &  13661 &$   1.8        $ & 0.016 &
$ 1230\pm 30$ & $   35\pm 0$ & C   \\
Z90735   &  86.917 & $+ 49.393$ &  18113 &$   2.3        $ & 0.015 &
$ 1180\pm 30$ & $   42\pm 0$ & C   \\
Q11002   &  86.983 & $+ 64.666$ &   8380 &$   2.5        $ & 0.010 &
$ 1270\pm 30$ & $   33\pm 0$ & --- \\
C03702   &  88.217 & $ -55.564$ &  71065 &$   0.9        $ & 0.052 &
$  420\pm 30$ & $   47\pm 0$ & MS  \\
S40518   &  91.373 & $+  1.134$ &  30450 &$   1.4        $ & 2.787 &
$  434\pm 22$ & $ 1552\pm 4$ & --- \\
Z90736   &  91.820 & $+ 30.217$ &  27775 &$   1.6        $ & 0.041 &
$  725\pm 10$ & $   61\pm 3$ & C   \\
M10310   &  94.027 & $+ 27.428$ &  12798 &$   2.3        $ & 0.043 &
$  712\pm  9$ & $  147\pm 3$ & --- \\
P20410   &  98.733 & $+ 29.775$ &  19748 &$   1.6        $ & 0.054 &
$  619\pm  8$ & $  115\pm 3$ & C   \\
B04604   &  99.293 & $+  3.738$ &  10076 &$   2.6        $ & 0.963 &
$  283\pm 16$ & $14237\pm 0$ & --- \\
S40524   & 100.514 & $+  8.623$ &  42076 &$   1.0        $ & 1.009 &
$  314\pm 16$ & $ 2574\pm 6$ & --- \\
S40564   & 100.608 & $ -13.070$ &   5080 &$   2.3        $ & 0.264 &
$  240\pm 18$ & $  320\pm 0$ & G   \\
S40543   & 100.787 & $+ 44.753$ &  10636 &$   2.2        $ & 0.032 &
$  970\pm 30$ & $   31\pm 0$ & C   \\
S40532   & 101.243 & $  -1.686$ &  18310 &$   1.7        $ & 0.537 &
$  252\pm 17$ & $ 1518\pm 5$ & G   \\
Z90711   & 101.294 & $+ 32.279$ &  21733 &$   1.8        $ & 0.034 &
$  688\pm 13$ & $   83\pm 3$ & C   \\
B01805   & 101.785 & $+ 59.630$ &   4164 &$   4          $ & 0.010 &
$ 1330\pm 40$ & $   35\pm 3$ & C   \\
Z90714   & 103.875 & $+ 36.157$ &  11051 &$   3          $ & 0.043 &
$  554\pm 14$ & $   88\pm 3$ & C   \\
A05101   & 104.063 & $+ 14.193$ &  14883 &$   1.7        $ &11.928 &
$  357\pm 20$ & $  361\pm 4$ & --- \\
S40522   & 105.720 & $  -5.122$ &   3969 &$   4          $ & 0.399 &
$  260\pm 30$ & $  897\pm 4$ & G   \\
A03406   & 110.865 & $+ 11.487$ &   1600 &$   5          $ & 0.702 &
$  460\pm 30$ & $  699\pm 4$ & --- \\
S40579   & 111.297 & $+ 33.579$ &  67002 &$   1.1        $ & 0.041 &
$  583\pm 22$ & $   56\pm 3$ & --- \\
S40510   & 117.187 & $+ 46.348$ &  28645 &$   1.3        $ & 0.012 &
$ 1090\pm 30$ & $    9\pm 4$ & C   \\
S40535   & 119.019 & $  -0.893$ &   3680 &$   4          $ & 1.696 &
$  410\pm 40$ & $ 2331\pm 0$ & H   \\
Z90709   & 123.725 & $+ 58.782$ &   8067 &$   4          $ & 0.011 &
$ 1170\pm 40$ & $   14\pm 2$ & C   \\
Z90712   & 125.299 & $+ 46.300$ &   5474 &$   4          $ & 0.018 &
$ 1000\pm 30$ & $   60\pm 3$ & C   \\
C02201   & 125.618 & $  -8.037$ &  17556 &$   2.4        $ & 0.416 &
$  440\pm 50$ & $  602\pm 4$ & H   \\
S40505   & 133.117 & $+ 55.658$ &  37331 &$   1.4        $ & 0.015 &
$  900\pm 50$ & $   19\pm 2$ & C   \\
S40529   & 133.963 & $  -4.989$ &  11836 &$   2.3        $ & 0.413 &
$  378\pm 22$ & $  443\pm 3$ & H   \\
S40562   & 134.391 & $+ 61.764$ &  14440 &$   1.6        $ & 0.015 &
$  850\pm 40$ & $   38\pm 2$ & C?  \\
S40570   & 138.060 & $ -11.120$ &  13778 &$   2.0        $ & 0.162 &
$  530\pm 30$ & $  410\pm 3$ & H   \\
S40580   & 141.384 & $+ 32.628$ &  15066 &$   1.8        $ & 0.028 &
$  610\pm 30$ & $   73\pm 3$ & A   \\
S40572   & 141.499 & $+  2.878$ &  13700 &$   2.1        $ & 1.318 &
$  287\pm 21$ & $ 1285\pm 0$ & H   \\
S40588   & 150.639 & $+ 58.933$ &  12401 &$   2.1        $ & 0.017 &
$ 1100\pm 50$ & $   16\pm 3$ & C?  \\
Z90727   & 151.433 & $+ 22.123$ &   2960 &$   4          $ & 0.092 &
$  440\pm 30$ & $  114\pm 3$ & --- \\
S40576   & 155.955 & $+  7.099$ &  29845 &$   1.4        $ & 0.593 &
$  320\pm 30$ & $  213\pm 3$ & --- \\
S40502   & 158.488 & $+  0.472$ &   5672 &$   3          $ & 1.174 &
$  330\pm 30$ & $ 1359\pm 4$ & --- \\
S40574   & 159.921 & $+ 58.496$ &   7510 &$   2.4        $ & 0.020 &
$  910\pm 30$ & $   40\pm 3$ & C   \\
S40587   & 161.916 & $+ 64.733$ &   8390 &$   2.3        $ & 0.014 &
$ 1210\pm 40$ & $   27\pm 3$ & M   \\
S40540   & 164.067 & $+ 49.004$ &  14779 &$   1.9        $ & 0.010 &
$ 1000\pm 40$ & $   66\pm 4$ & A   \\
S40541   & 164.817 & $+ 36.961$ &   2290 &$   4          $ & 0.038 &
$  610\pm 50$ & $   54\pm 0$ & A   \\
Z90732   & 165.287 & $+ 61.408$ &  10347 &$   4          $ & 0.012 &
$ 1420\pm 40$ & $   36\pm 3$ & --- \\
P10414   & 166.158 & $+ 10.475$ &   7960 &$   3          $ & 0.234 &
$  330\pm 30$ & $  365\pm 3$ & --- \\
S40513   & 178.879 & $+ 59.009$ &  12260 &$   2.4        $ & 0.014 &
$ 1500\pm 50$ & $   31\pm 3$ & M   \\
P10409   & 180.973 & $ -20.249$ &  11883 &$   2.3        $ & 0.587 &
$  450\pm 30$ & $  813\pm 0$ &AnCen\\
S40531   & 188.956 & $ -40.098$ &   5910 &$   3          $ & 0.188 &
$  710\pm 30$ & $ 1402\pm 4$ & --- \\
P10407   & 195.846 & $ -48.051$ &  27675 &$   1.3        $ & 0.039 &
$ 1490\pm 60$ & $  476\pm 0$ & --- \\
S40504   & 197.880 & $+ 17.400$ &   1635 &$   5          $ & 0.051 &
$ 1680\pm 50$ & $  397\pm 3$ & --- \\
A03407   & 243.721 & $ -26.124$ &   8150 &$   3          $ & 0.036 &
$  770\pm 30$ & \nodata      & --- \\
S40509   & 247.554 & $+ 47.752$ &   5721 &$   3          $ & 0.076 &
$  550\pm 30$ & $  228\pm 4$ & WA? \\
S40573   & 264.191 & $ -34.510$ &  21100 &$   1.8        $ & 0.040 &
$  620\pm 40$ & \nodata      & W?  \\
Z90726   & 266.478 & $ -29.397$ &  21110 &$   1.7        $ & 0.046 &
$  504\pm 16$ & \nodata      & --- \\
A04802   & 267.366 & $  -7.476$ &   6300 &$   4          $ & 0.965 &
$  555\pm 22$ & \nodata      & --- \\
S40582   & 270.079 & $ -30.615$ &  21403 &$   1.8        $ & 0.052 &
$  451\pm 15$ & \nodata      & --- \\
C05501   & 276.101 & $ -33.239$ &   9470 &$   3          $ & 0.075 &
$  573\pm 14$ & \nodata      & MS  \\
Z90719   & 276.177 & $ -47.638$ &  10390 &$   3          $ & 0.074 &
$  780\pm 40$ & \nodata      & MS  \\
A11102   & 277.764 & $ -33.014$ &   4240 &$   6          $ & 0.075 &
$  599\pm 12$ & \nodata      & MS  \\
A11101   & 277.767 & $ -33.023$ &   4230 &$   6          $ & 0.075 &
$  599\pm 12$ & \nodata      & MS  \\
A04905   & 278.649 & $ -33.170$ &   2738 &$   5          $ & 0.075 &
$  537\pm 11$ & \nodata      & MS  \\
S40593   & 278.892 & $ -36.323$ &   5561 &$   4          $ & 0.075 &
$  522\pm 16$ & \nodata      & MS  \\
B09401   & 280.309 & $ -32.795$ &   4200 &$   5          $ & 0.075 &
$  701\pm 20$ & \nodata      & MS  \\
P20421   & 281.621 & $ -23.493$ &   4582 &$   4          $ & 0.224 &
$  660\pm 30$ & \nodata      & MS  \\
S51303   & 286.148 & $+  0.957$ &   3444 &$   5          $ & 1.088 &
$  360\pm 40$ & \nodata      & --- \\
S40553   & 290.948 & $  -0.488$ &  11878 &$   2.5        $ & 1.412 &
$  300\pm 30$ & \nodata      & --- \\
S40507   & 292.316 & $  -4.829$ &  14362 &$   3          $ & 0.800 &
$  290\pm 23$ & \nodata      & --- \\
S40558   & 295.613 & $  -0.240$ &   7390 &$   3          $ & 3.527 &
$  430\pm 30$ & \nodata      & WD? \\
S51401   & 299.156 & $ -10.943$ &  11553 &$   3          $ & 0.238 &
$  190\pm 30$ & \nodata      & WE? \\
A01002   & 299.854 & $ -30.679$ &   2030 &$   4          $ & 0.080 &
$  650\pm 30$ & \nodata      & --- \\
B06801   & 300.171 & $+  5.650$ &  12110 &$   3          $ & 0.835 &
$  360\pm 40$ & \nodata      & WD? \\
S40589   & 318.630 & $ -29.168$ &   9854 &$   2.1        $ & 0.115 &
$  470\pm 60$ & \nodata      & --- \\
A10001   & 321.537 & $ -15.259$ &   9412 &$   2.3        $ & 0.102 &
$  480\pm 60$ & \nodata      & WE  \\
Z90708   & 322.827 & $ -58.853$ &  12420 &$   4          $ & 0.011 &
$ 1250\pm 60$ & \nodata      & MS  \\
Z90737   & 327.712 & $ -23.823$ &  29401 &$   1.5        $ & 0.084 &
$  430\pm 70$ & \nodata      & --- \\
S40563   & 329.883 & $  -7.017$ &  38174 &$   1.5        $ & 0.355 &
$  460\pm 40$ & \nodata      & --- \\
S40569   & 330.353 & $  -6.828$ &  10580 &$   1.7        $ & 0.287 &
$  480\pm 40$ & \nodata      & --- \\
C07604   & 336.624 & $ -25.642$ &  14454 &$   2.3        $ & 0.059 &
$  680\pm 80$ & \nodata      & --- \\
P20423   & 339.728 & $ -48.055$ &   4405 &$   4          $ & 0.025 &
$ 2030\pm 60$ & \nodata      & --- \\
S40526   & 343.165 & $  -4.762$ &   7058 &$   2.1        $ & 0.554 &
$  330\pm 30$ & \nodata      & --- \\
C02301   & 346.268 & $ -57.954$ &   5700 &$   3          $ & 0.009 &
$ 1190\pm 50$ & \nodata      & --- \\
S40514   & 349.573 & $  -9.088$ &  11022 &$   3          $ & 0.181 &
$  280\pm 70$ & \nodata      & --- \\
Z90703   & 351.251 & $ -58.997$ &  10750 &$   4          $ & 0.011 &
$  970\pm 50$ & \nodata      & --- \\
Z90702   & 354.668 & $ -49.893$ &  15290 &$   1.8        $ & 0.017 &
$  960\pm 50$ & \nodata      & --- \\
P20419   & 358.794 & $+ 24.180$ &  14122 &$   2.3        $ & 0.311 &
$  690\pm100$ & $  812\pm 4$ & --- \\
\enddata
\tablenotetext{a}{Exposure time is night only.}
\tablenotetext{b}{1\,LU = 1~photon~s$^{-1}$\,cm$^{-2}$\,sr$^{-1}$; for
\ion{O}{6} $\lambda$1032, 1\,LU =
$4.5\times10^{-22}$~erg~s$^{-1}$\,cm$^{-2}$\,arcsec$^{-2}$.}
\tablenotetext{c}{Total reddening \citep{sfd}.}
\tablenotetext{d}{{\em ROSAT} $\onequarter$ keV emission with 1\,RU =
$10^{-6}$~counts~s$^{-1}$\,arcmin$^{-2}$ \citep{snow}.}
\tablenotetext{e}{H$\alpha$ intensity integrated over velocity range
$-80$\,km\,s$^{-1}<v<+80$\,km\,s$^{-1}$; an uncertainty of 0 means that the 
intensity is the average of the neighboring pointings \citep{haf}; for
H$\alpha$, 1\,LU = $7.1\times10^{-23}$~erg~s$^{-1}$\,cm$^{-2}$\,arcsec$^{-2}$.}
\tablenotetext{f}{Based on \ion{H}{1} map in \citet{wak03}.}
\end{deluxetable}

\clearpage

\begin{deluxetable}{lrrcccccccc}
\tabletypesize{\footnotesize}
\rotate
\tablewidth{0pt}
\tablecaption{\label{pub} PUBLISHED \ion{O}{6} $\lambda$1032 MEASUREMENTS AND
2$\sigma$ UPPER LIMITS}
\tablehead{\colhead{Target} & \colhead{$l$} & \colhead{$b$} &
\colhead{$I_{1032}$\tablenotemark{a}} & \colhead{FWHM} & \colhead{$v_{\rm LSR}$}
& \colhead{} & \colhead{SXR\tablenotemark{c}} & \colhead{$I_{{\rm
H}\alpha}$\tablenotemark{d}} & \colhead{} & \colhead{} \\
\colhead{ID} & \colhead{(deg.)} & \colhead{(deg.)} & \colhead{(10$^3$\,LU)} &
\colhead{(km\,s$^{-1}$)} & \colhead{(km\,s$^{-1}$)} &
\colhead{$E(B-V)$\tablenotemark{b}} & \colhead{(RU)} & \colhead{(10$^3$\,LU)} &
\colhead{HVC\tablenotemark{e}} & \colhead{Ref.}}
\startdata
SL1      &  57.6   & $+ 88.0  $ & $   2.0\pm  0.6$ & $ 23  \pm 55  $ &
 $ -17  \pm 9  $ & 0.008 & $ 4430\pm100$ & $   16\pm 3$ & --- & 1 \\
SL2      & 284.2   & $+ 74.5  $ & $   2.9\pm  0.7$ & $ 80          $ &
 $  84  \pm15  $ & 0.023 & $ 5530\pm120$ & $   37\pm 3$ & --- & 1 \\
SL3      & 315.0   & $ -41.3  $ & $   2.9\pm  0.3$ & $160          $ &
 $  64         $ & 0.036 & $  930\pm 60$ & \nodata      & --- & 2 \\
SL4      & 113.0   & $+ 70.7  $ & $   2.6\pm  0.4$ & $ 75          $ &
 $  10         $ & 0.008 & $ 1000\pm 40$ & $   37\pm 2$ & --- & 3 \\
SL5      & 162.7   & $+ 57.0  $ & $   2.5\pm  0.7$ & $150          $ &
 $ -16  \pm22  $ & 0.013 & $ 1130\pm 50$ & $   15\pm 3$ & C   & 4 \\
SL6      & 156.3   & $+ 57.8  $ & $   3.3\pm  1.1$ & $210          $ &
 $ -50  \pm30  $ & 0.011 & $ 1140\pm 70$ & \nodata      & C   & 4 \\
SL7      &  95.4   & $+ 36.1  $ & $   1.7\pm  0.4$ & $ 75  \pm  3  $ &
 $ -50  \pm30  $ & 0.024 & $  944\pm 15$ & $   81\pm 0$ & C   & 5 \\
SL8      &  99.3   & $+ 43.3  $ &$<   1.7        $ & \nodata         &
 \nodata         & 0.034 & $ 1160\pm 30$ & $   37\pm 3$ & C   & 5 \\
SL9      & 278.6   & $ -45.3  $ &$<   0.5        $ & \nodata         &
 \nodata         & 0.153 & $  600\pm 40$ & \nodata      & (MS) & 6 \\
SL10     & 348.1   & $ -66.3  $ & $   2.2\pm  0.5$ & $ 60  \pm 40  $ &
 $  -5  \pm 9  $ & 0.021 & $  870\pm 70$ & \nodata      & --- & 7 \\
SL11     & 346.6   & $ -39.5  $ & $   6.1\pm  1.2$ & $140  \pm 40  $ &
 $ -28  \pm15  $ & 0.035 & $  790\pm 50$ & \nodata      & --- & 7 \\
\enddata
\tablerefs{(1) Dixon et al. 2001; (2) Shelton et al. 2001; (3) Shelton 2002; (4)
Welsh et al. 2002; (5) Otte et al. 2003a; (6) Shelton 2003; (7) Otte et al.
2003b}
\tablenotetext{a}{1\,LU = 1~photon~s$^{-1}$\,cm$^{-2}$\,sr$^{-1}$; for
\ion{O}{6} $\lambda$1032, 1\,LU =
$4.5\times10^{-22}$~erg~s$^{-1}$\,cm$^{-2}$\,arcsec$^{-2}$.}
\tablenotetext{b}{Total reddening \citep{sfd}.}
\tablenotetext{c}{{\em ROSAT} $\onequarter$ keV emission with 1\,RU =
$10^{-6}$~counts~s$^{-1}$\,arcmin$^{-2}$ \citep{snow}.}
\tablenotetext{d}{H$\alpha$ intensity integrated over velocity range
$-80$\,km\,s$^{-1}<v<+80$\,km\,s$^{-1}$; an uncertainty of 0 means that the 
intensity is the average of the neighboring pointings \citep{haf}; for
H$\alpha$, 1\,LU = $7.1\times10^{-23}$~erg~s$^{-1}$\,cm$^{-2}$\,arcsec$^{-2}$.}
\tablenotetext{e}{Based on \ion{H}{1} map in \citet{wak03}.}
\end{deluxetable}

\clearpage

\begin{deluxetable}{lcrrl}
\tablewidth{0pt}
\tablecaption{\label{locs} DETECTIONS: H$\alpha$ MORPHOLOGY AND LOCATION OF
\ion{O}{6} GAS}
\tablehead{\colhead{Target} & \colhead{H$\alpha$} & \colhead{$|z|$} &
\colhead{$D$} & \colhead{} \\
\colhead{ID} & \colhead{Morphology} & \colhead{(kpc)} & \colhead{(kpc)} &
\colhead{Comment\tablenotemark{a}}}
\startdata
S40523 & diffuse & 0--1.2 & 0--7.7 & near disk \\
S40521 & filament/blob & \nodata & \nodata & not corotating \\
S40560 & filament & $1.4^{+2.0}_{-1.4}$ & $2.4^{+3.7}_{-2.4}$ & near disk
 \\[1ex]
Z90715 & filament & $1.0^{+2.6}_{-1.0}$ & $1.8^{+4.4}_{-1.8}$ & near disk \\
S40581 & diffuse & \nodata & \nodata & not corotating \\
P20411 & diffuse & $1.5^{+0.7}_{-0.7}$ & $3.1^{+1.5}_{-1.4}$ & near disk \\
S30402 & \ion{H}{2} & \nodata & \nodata & not corotating \\
P10413 & blob & $0.6^{+?}_{-0.6}$ & $7.4^{+?}_{-7.4}$ & near disk \\
S40508 & diffuse & \nodata & \nodata & not corotating \\
P12012 & filament & 0--0.5 & 0--2.1 & if corotating, probably local \\
P10418 & filament & \nodata & \nodata & not corotating \\
P10411 & filament/bubble & 0--0.1 & 0--0.2 & local \\
S40537 & filament/bubble & \nodata & \nodata & not corotating \\
P20517 & bubble & \nodata & \nodata & not corotating \\
S40506 & bubble & 0--0.7 & 0--3.6 & if corotating, local or near disk \\
S40591 & filament/bubble & 0--0.01 & 0--1.5 & local \\[1ex]
 \nodata & \nodata & $0.05^{+0.01}_{-0.01}$ & $8.2^{+1.6}_{-1.8}$ & \nodata
 \\[1ex]
S40557 & diffuse & $1.2^{+0.2}_{-0.3}$ & $5.3^{+0.7}_{-1.5}$ & disk \\[1ex]
 \nodata & \nodata & $1.6^{+0.3}_{-0.2}$ & $6.7^{+1.5}_{-0.7}$ & \nodata \\[1ex]
A04604 & diffuse & $0.2^{+1.0}_{-0.2}$ & $0.4^{+2.0}_{-0.4}$ & local or near
 disk \\[1ex]
 \nodata & \nodata & $6.7^{+1.2}_{-1.0}$ & $13.5^{+2.3}_{-2.1}$ & \nodata
 \\[1ex]
P10425 & \ion{H}{2}/blob & 0.0 & $0.3^{+1.2}_{-0.3}$ & local \\[1ex]
 \nodata & \nodata & $0.1^{+0.1}_{-0}$ & $11.9^{+1.2}_{-1.2}$ & \nodata \\[1ex]
S40590 & diffuse & $0.4^{+0.2}_{-0.1}$ & $2.6^{+0.9}_{-0.8}$ & near disk \\[1ex]
 \nodata & \nodata & $1.7^{+0.2}_{-0.1}$ & $10.5^{+0.8}_{-0.8}$ & \nodata
 \\[1ex]
C07602 & diffuse & $0.5^{+0.7}_{-0.5}$ & $1.2^{+1.6}_{-1.2}$ & near disk \\[1ex]
 \nodata & \nodata & $7.0^{+1.0}_{-0.7}$ & $16.1^{+2.3}_{-1.6}$ & \nodata
 \\[1ex]
P20422 & diffuse & $1.2^{+0.5}_{-0.7}$ & $2.1^{+1.1}_{-1.3}$ & nead disk \\[1ex]
 \nodata & \nodata & $8.9^{+0.7}_{-0.5}$ & $16.5^{+1.2}_{-1.1}$ & \nodata
 \\[1ex]
S40568 & diffuse & 0--3.2 & 0--6.8 & local or near disk \\
 \nodata & \nodata & 5.8--? & 12.3--? & \nodata \\
\enddata
\tablenotetext{a}{{Likely} location; see text for more information.}
\end{deluxetable}

\clearpage

\clearpage
\epsscale{1.0}
\plotone{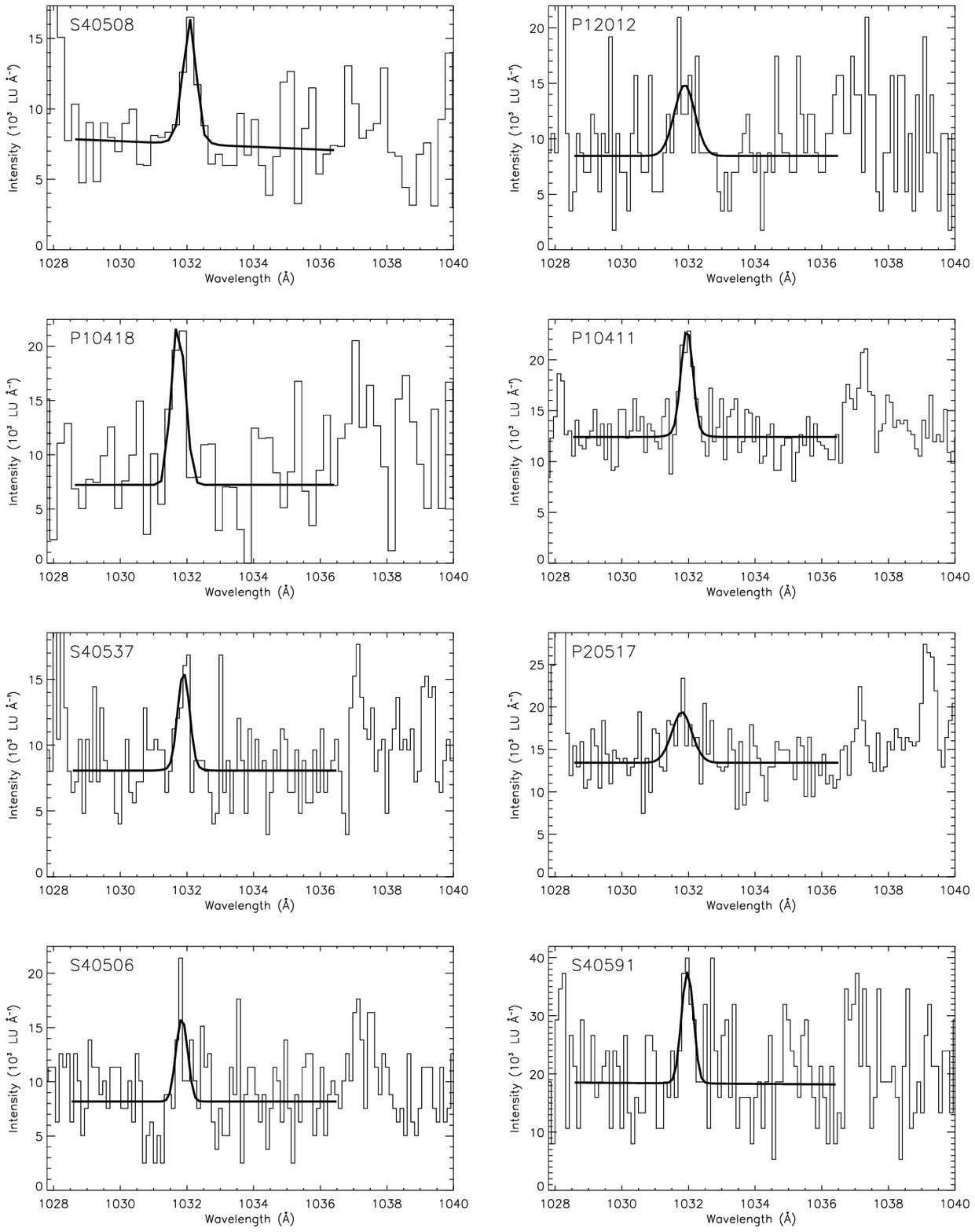}
\centerline{Fig. 3. --- Continued.}
\clearpage
\plotone{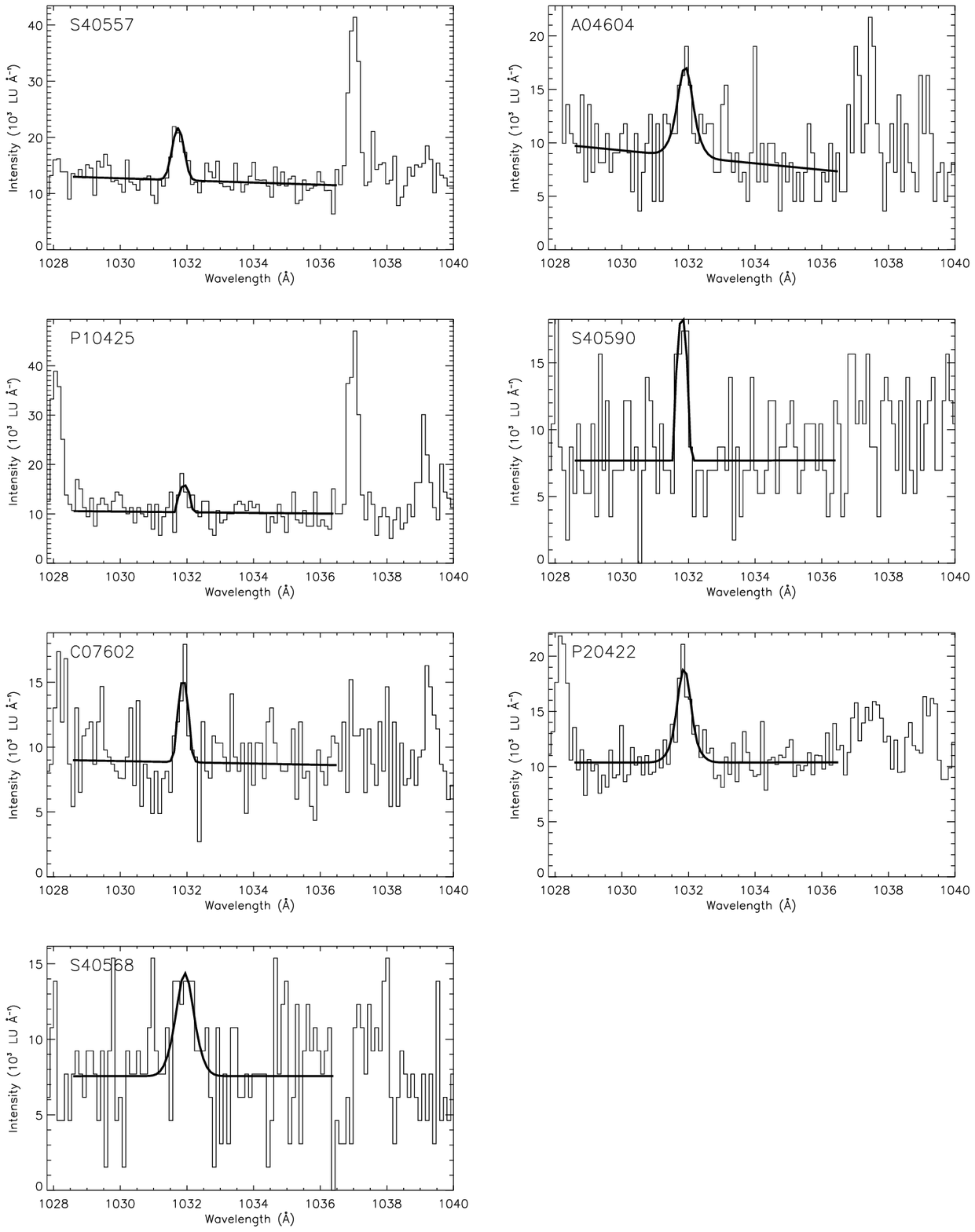}
\centerline{Fig. 3. --- Continued.}
\clearpage










\clearpage

\begin{thebibliography}{}
\bibitem[Bannister et al.(2003)]{bann} Bannister, N. P., Barstow, M. A., 
Holberg, J. B., \& Bruhweiler, F. C. 2003, \mnras, 341, 477
\bibitem[Bowen et al.(2006)]{bowen} Bowen, D. V., Jenkins, E. B., Tripp, T. M.,
Sembach, K. R., \& Savage, B. D. 2006, in ASP Conf. Ser. 348, Astrophysics in
the Far Ultraviolet: Five Years of Discovery with FUSE, ed. G. Sonneborn, H. W.
Moos, \& B.-G. Andersson (San Francisco: ASP), 412
\bibitem[Collins et al.(2005)]{col05} Collins, J. A., Shull, J. M., \& Giroux,
M. L. 2005, \apj, 623, 196
\bibitem[Dixon et al.(2001)]{dixon} Dixon, W. V., Sallmen, S., Hurwitz, M., \&
Lieu, R. 2001, \apjl, 552, L69
\bibitem[Dixon et al.(2006)]{dix06} Dixon, W. V., Sankrit, R., \& Otte, B. 2006,
\apjs, in press
\bibitem[Fitzpatrick (1999)]{fitz} Fitzpatrick, E. L. 1999, \pasp, 111, 63
\bibitem[Fleming et al.(1996)]{fle} Fleming, T. A., Snowden, S. L., Pfeffermann,
E., Briel, U., \& Greiner, J. 1996, \aap, 316, 147
\bibitem[Gaustad et al.(2001)]{gau} Gaustad, J. E., McCullough, P. R., Rosing,
W., \& Van Buren, D. 2001, \pasp, 113, 1326
\bibitem[Haffner et al.(2003)]{haf} Haffner, L. M., Reynolds, R. J., Tufte, S.
L., Madsen, G. J., Jaehnig, K. P., \& Percival, J. W. 2003, \apjs, 149, 405
\bibitem[Holberg et al.(1998)]{holb} Holberg, J. B., Barstow, M. A., \& Sion, E.
M. 1998, \apjs, 119, 207
\bibitem[Howk et al.(2002)]{howk} Howk, J. C., Savage, B. D., Sembach, K. R., \&
Hoopes, C. G., 2002, \apj, 572, 264
\bibitem[Indebetouw \& Shull(2004a)]{inda} Indebetouw, R., \& Shull, J. M.
2004a, \apj, 605, 205
\bibitem[Indebetouw \& Shull(2004b)]{indb} Indebetouw, R., \& Shull, J. M.
2004b, \apj, 607, 309
\bibitem[Jenkins (1978a)]{j78a} Jenkins, E. B. 1978a, \apj, 219, 845
\bibitem[Jenkins (1978b)]{j78b} Jenkins, E. B. 1978b, \apj, 220, 107
\bibitem[Kriss (1994)]{kriss} Kriss, G. A. 1994, in ASP Conf. Ser. 61,
Astronomical Data Analysis Software and Systems III, ed. D. R. Crabtree, R. J.
Hanisch, \& J. Barnes (San Francisco: ASP), 437
\bibitem[Landini \& Monsignori Fossi(1990)]{lan} Landini, M, \& Monsignori
Fossi, B. C. 1990, A\&AS, 82, 229
\bibitem[Moos et al.(2000)]{moos} Moos, H. W., et al. 2000, \apjl, 538, L1
\bibitem[Otte et al.(2003a)]{o03} Otte, B., Dixon, W. V., \& Sankrit, R. 2003a,
\apjl, 586, L53
\bibitem[Otte et al.(2003b)]{o03b} Otte, B., Dixon, W. V., \& Sankrit, R. 2003b,
AAS, 203, 110.12
\bibitem[Otte et al.(2004)]{o04} Otte, B., Dixon, W. V., \& Sankrit, R. 2004,
\apjl, 606, L143
\bibitem[Sahnow et al.(2000)]{sahn} Sahnow, D. J., et al. 2000, \apjl, 538, L7
\bibitem[Savage \& Lehner (2006)]{sav06} Savage, B. D., \& Lehner, N. 2006,
\apjs, 162, 134
\bibitem[Savage et al.(2003)]{sav} Savage, B. D., et al. 2003, \apjs, 146, 125
\bibitem[Schlegel et al.(1998)]{sfd} Schlegel, D. J., Finkbeiner, D. P., \&
Davis, M. 1998, \apj, 500, 525
\bibitem[Sembach \& Savage(1992)]{sem92} Sembach, K. R., \& Savage, B. D. 1992,
\apjs, 83, 147
\bibitem[Sembach(1993)]{sem93} Sembach, K. R. 1993, \pasp, 105, 983
\bibitem[Sembach et al.(1995)]{sem95} Sembach, K. R., Savage, B. D., Lu, L., \&
Murphy, E. M. 1995, \apj, 451, 616
\bibitem[Sembach et al.(1999)]{sem99} Sembach, K. R., Savage, B. D., Lu, L., \&
Murphy, E. M. 1999, \apj, 515, 108
\bibitem[Sembach et al.(2003)]{sem} Sembach, K. R., et al. 2003, \apjs, 146, 165
\bibitem[Shelton et al.(2001)]{sh01} Shelton, R. L., et al. 2001, \apj, 560, 730
\bibitem[Shelton(2002)]{sh02} Shelton, R. L. 2002, \apj, 569, 758
\bibitem[Shelton(2003)]{sh03} Shelton, R. L. 2003, \apj, 589, 261
\bibitem[Shull \& Slavin(1994)]{ss} Shull, J. M., \& Slavin, J. D. 1994, \apj,
427, 784
\bibitem[Shull et al.(2000)]{sh00} Shull, J. M., et al. 2000, \apj, 538, L73
\bibitem[Slavin et al.(1993)]{sla} Slavin, J. D., Shull, J. M., \& Begelman, M.
C. 1993, \apj, 407, 83
\bibitem[Snowden et al.(1997)]{snow} Snowden, S. L., et al. 1997, \apj, 485, 125
\bibitem[Spitzer (1996)]{spi} Spitzer, L., Jr. 1996, \apj, 458, L29
\bibitem[Sutherland \& Dopita(1993)]{sd} Sutherland, R. S., \& Dopita, M. A.
1993, \apjs, 88, 253
\bibitem[Wakker et al.(2003)]{wak03} Wakker, B. P. et al. 2003, \apjs, 146, 1
\bibitem[Welsh et al.(2002)]{wel} Welsh, B. Y., Sallmen, S., Sfeir, D.,
Shelton, R. L., \& Lallement, R. 2002, \aap, 394, 691
\bibitem[Young et al.(2001)]{young} Young, P. R., Dupree, A. K., Wood, B. E.,
Redfield, S., Linsky, J. L., Ake, T. B., \& Moos, H. W., \apjl, 555, L121
\end{thebibliography}
\end{document}